\documentclass[11pt]{myArticle}       

\usepackage{amsmath,amssymb,amsfonts} 
\usepackage{graphicx}
\usepackage{subfig}
\usepackage{epsfig}
\usepackage{color}                    
\usepackage{hyperref}                 
\usepackage{myFancyhdr}               
\usepackage{makeidx}                  
\usepackage{helvet}                   
\usepackage{setspace}                 
\usepackage{fancybox}                 
\usepackage{float}                    
\usepackage{xspace}                   
\usepackage{wasysym}                  
\usepackage[numbers,sort&compress]{natbib}
\usepackage{natbib}
\usepackage{hypernat}
\usepackage{wrapfig}
\usepackage{lineno}

\definecolor{mygray}{rgb}{0.3,0.32,0.35}
\definecolor{darkblue1}{rgb}{0,0,.2}
\definecolor{darkblue}{rgb}{0,0,.3}
\definecolor{darkred}{rgb}{0.5,0,0}
\pagecolor{white} 
\color{black}     
\hypersetup{breaklinks=true, 
            colorlinks=true, 
            linkcolor=darkblue1, 
            menucolor=darkblue1, 
            urlcolor=darkblue1,
            citecolor=darkblue1,
            pdftitle={The EWK Fit},
            pdfauthor={Gfitter Group},
            pdfsubject={The EWK Fit},
            pdfkeywords={},
            pdfproducer={Gfitter Group}
}
\newcommand\defaultSingleFigureScale{0.70}
\newcommand\defaultDoubleFigureScale{0.495}

\bibstyle{plain}
\renewcommand{\headrulewidth}{0.5pt}    
\renewcommand{\footrulewidth}{0.5pt}    

\newcommand\allFontSize{\small}

\newenvironment{myquote}
               {\list{}{\leftmargin0cm}%
                \item\relax}
               {\endlist}

\newcommand\detailsSize{\allFontSize}
\newenvironment{details}%
{\begin{myquote}\vspace{-0.2cm}\detailsSize}{\end{myquote}\vspace{-0.2cm}}

\newfloat{codeexample}{H}{loc}
\floatname{codeexample}{\sf\allFontSize Code Example}

\newlength{\gfitterboxwidth}
\definecolor{DarkGray}{rgb}{0.4,0.42,0.45}
\definecolor{LightGray}{rgb}{0.97,0.98,0.98}

{\VerbatimEnvironment\normalsize
\setlength{\gfitterboxwidth}{\textwidth}\addtolength{\gfitterboxwidth}{-2.3\fboxsep}%
\begin{Sbox}\begin{minipage}[t]{\gfitterboxwidth}\begin{Verbatim}}%
{\end{Verbatim}\end{minipage}\end{Sbox}\vspace{1ex} \fcolorbox{DarkGray}{LightGray}{\TheSbox}}

\newfloat{option}{H}{loc}
\floatname{option}{\sf\allFontSize Option Table}

\usepackage{tabularx}
{\setlength{\gfitterboxwidth}{\textwidth}\addtolength{\gfitterboxwidth}{-2.3\fboxsep}%
\begin{Sbox}\begin{tabular*}{\gfitterboxwidth}{>{\rule[-0.5ex]{.0ex}{3.0ex}\tt\small}p{3cm}>{\tt\small}p{4.5cm}>{\small}p{5.7cm}}
&&\\[-0.7cm]
\rm\small Option  & \rm\small Values    & \rm\small Description\\[0.1cm]\hline
&&\\[-0.45cm]}%
{\\[-0.1cm]\end{tabular*}\end{Sbox}\vspace{1ex} \fbox{\TheSbox}}

{\setlength{\gfitterboxwidth}{\textwidth}\addtolength{\gfitterboxwidth}{-2.3\fboxsep}%
\begin{Sbox}\begin{tabular*}{\gfitterboxwidth}{>{\rule[-0.5ex]{.0ex}{3.0ex}\tt\small}p{3cm}>{\tt\small}p{3.5cm}>{\small}p{6.7cm}}
&&\\[-0.7cm]
\rm\small Option  & \rm\small Values    & \rm\small Description\\[0.1cm]\hline
&&\\[-0.45cm]}%
{\\[-0.1cm]\end{tabular*}\end{Sbox}\vspace{1ex} \fbox{\TheSbox}}

{\setlength{\gfitterboxwidth}{\textwidth}\addtolength{\gfitterboxwidth}{-2.3\fboxsep}%
\begin{Sbox}\begin{tabular*}{\gfitterboxwidth}{>{\rule[-0.5ex]{.0ex}{3.5ex}\tt\small}p{3.0cm}>{\tt\small}p{3.0cm}>{\small}p{7.2cm}}
&&\\[-0.7cm]
\rm\small Option  & \rm\small Values    & \rm\small Description\\[0.1cm]\hline
&&\\[-0.45cm]}%
{\\[-0.1cm]\end{tabular*}\end{Sbox}\vspace{1ex} \fbox{\TheSbox}}

{\setlength{\gfitterboxwidth}{\textwidth}\addtolength{\gfitterboxwidth}{-2.3\fboxsep}%
\begin{Sbox}\begin{tabular*}{\gfitterboxwidth}{>{\rule[-0.5ex]{.0ex}{3.0ex}\tt\small}p{2.5cm}>{\tt\small}p{3.5cm}>{\small}p{7.2cm}}
&&\\[-0.7cm]
\rm\small Option  & \rm\small Values    & \rm\small Description\\[0.1cm]\hline
&&\\[-0.45cm]}%
{\\[-0.1cm]\end{tabular*}\end{Sbox}\vspace{1ex} \fbox{\TheSbox}}

{\setlength{\gfitterboxwidth}{\textwidth}\addtolength{\gfitterboxwidth}{-2\fboxsep}%
\begin{Sbox}\begin{tabular*}{\gfitterboxwidth}{>{\rule[-0.5ex]{.0ex}{3.0ex}\tt\small}p{4.5cm}>{\tt\small}p{3.2cm}>{\small}p{5.5cm}}
&&\\[-0.7cm]
\rm\small Option  & \rm\small Values    & \rm\small Description\\[0.1cm]\hline
&&\\[-0.45cm]}%
{\\[-0.1cm]\end{tabular*}\end{Sbox}\vspace{1ex} \fbox{\TheSbox}}

{\setlength{\gfitterboxwidth}{\textwidth}\addtolength{\gfitterboxwidth}{-2\fboxsep}%
\begin{Sbox}\begin{tabular*}{\gfitterboxwidth}{>{\rule[-0.5ex]{.0ex}{3.0ex}\tt\small}p{4.0cm}>{\tt\small}p{3.0cm}>{\small}p{6.2cm}}
&&\\[-0.7cm]
\rm\small Option  & \rm\small Values    & \rm\small Description\\[0.1cm]\hline
&&\\[-0.45cm]}%
{\\[-0.1cm]\end{tabular*}\end{Sbox}\vspace{1ex} \fbox{\TheSbox}}

\newfloat{programs}{H}{loc}
\floatname{programs}{\sf Table}

\usepackage{tabularx}
{\setlength{\gfitterboxwidth}{\textwidth}\addtolength{\gfitterboxwidth}{-2\fboxsep}%
\begin{Sbox}\begin{tabular*}{\gfitterboxwidth}{>{\rule[-0.5ex]{.0ex}{3.0ex}\tt\small}p{4.1cm}>{\small}p{9.5cm}}
&\\[-0.7cm]
\rm\small Macro & \rm\small Description\\[0.1cm]\hline
&\\[-0.45cm]}%
{\\[-0.1cm]\end{tabular*}\end{Sbox}\vspace{1ex} \fbox{\TheSbox}}

\newcommand{\MSbar}{\ensuremath{\overline{\rm MS}}\xspace}
\mathchardef\Upsilon="7107
\def\Y#1S{\ensuremath{\Upsilon{(#1S)}}\xspace}

\newcommand{\mt}{\ensuremath{m_{t}}\xspace}

\newcommand{\MH}{\ensuremath{M_{H}}\xspace}
\newcommand{\as}{\ensuremath{\alpha_{\scriptscriptstyle S}}\xspace}

\renewcommand\l{\ell}

\newcommand{\Kbar    }{\kern 0.2em\overline{\kern -0.2em K}{}\xspace}

\newcommand{\Kz      }{\ensuremath{K^0}\xspace}
\newcommand{\Kzb     }{\ensuremath{\Kbar^0}\xspace}
\newcommand{\KzKzb   }{\ensuremath{\Kz \kern -0.16em \Kzb}\xspace}
\newcommand{\Kp      }{\ensuremath{K^+}\xspace}
\newcommand{\Km      }{\ensuremath{K^-}\xspace}

\newcommand{\KpKm    }{\ensuremath{\Kp \kern -0.16em \Km}\xspace}

\newcommand\Dbar    {\kern 0.18em\overline{\kern -0.18em D}{}\xspace}

\newcommand\Bbar    {\kern 0.18em\overline{\kern -0.18em B}{}\xspace}

\newcommand\Bz      {\ensuremath{B^0}\xspace}

\newcommand\Bzb     {\ensuremath{\Bbar^0}\xspace}
\newcommand\Bu      {\ensuremath{B^+}\xspace}
\newcommand\Bub     {\ensuremath{B^-}\xspace}

\newcommand\BpBm    {\ensuremath{\Bu {\kern -0.16em \Bub}}\xspace}
\newcommand\Bs      {\ensuremath{B^0_{s}}\xspace}
\newcommand\Bsb     {\ensuremath{\Bbar^0_{s}}\xspace}

\newcommand\BzBzb   {\ensuremath{\Bz {\kern -0.16em \Bzb}}\xspace}
\newcommand\BszBszb {\ensuremath{\Bs {\kern -0.16em \Bsb}}\xspace}

\newcommand{\ft}{\footnotesize}
\newcommand{\multic}{\multicolumn}

\newcommand{\nub}{\ensuremath{\overline{\nu}}\xspace}

\newcommand{\BR}{\ensuremath{{\mathcal B}}\xspace}

\newcommand{\tev}{\ensuremath{\mathrm{Te\kern -0.1em V}}\xspace}
\newcommand{\gev}{\ensuremath{\mathrm{Ge\kern -0.1em V}}\xspace}
\newcommand{\mev}{\ensuremath{\mathrm{Me\kern -0.1em V}}\xspace}
\newcommand{\kev}{\ensuremath{\mathrm{ke\kern -0.1em V}}\xspace}
\newcommand{\ev}{\ensuremath{\mathrm{e\kern -0.1em V}}\xspace}
\newcommand{\gevc}{\ensuremath{{\mathrm{Ge\kern -0.1em V\!/}c}}\xspace}
\newcommand{\mevc}{\ensuremath{{\mathrm{Me\kern -0.1em V\!/}c}}\xspace}
\newcommand{\gevcc}{\ensuremath{{\mathrm{Ge\kern -0.1em V\!/}c^2}}\xspace}
\newcommand{\mevcc}{\ensuremath{{\mathrm{Me\kern -0.1em V\!/}c^2}}\xspace}

\newcommand{\bei}{\begin{itemize}}
\newcommand{\eei}{\end{itemize}}
\newcommand{\beq}{\begin{equation}}
\newcommand{\eeq}{\end{equation}}
\newcommand{\beqn}{\begin{eqnarray}}
\newcommand{\eeqn}{\end{eqnarray}}
\newcommand{\beqns}{\begin{eqnarray*}}
\newcommand{\eeqns}{\end{eqnarray*}}
\newcommand{\bitm}{\begin{itemize}}
\newcommand{\eitm}{\end{itemize}}

\newcommand{\dahadZf}{\ensuremath{\Delta\alpha_{\rm had}^{(5)}(M_Z^2)}\xspace}
\newcommand{\dalphaHadMZ}{\ensuremath{\Delta\alpha_{\rm had}^{(5)}(M_Z^2)}\xspace}

\newcommand\rs{\raisebox{1.5ex}[-1.5ex]}

\def\@citex[#1]#2{\if@filesw\immediate\write\@auxout{\string\citation{#2}}\fi
  \@tempcnta\z@\@tempcntb\m@ne\def\@citea{}\@cite{\@for\@citeb:=#2\do
    {\@ifundefined
       {b@\@citeb}{\@citeo\@tempcntb\m@ne\@citea
        \def\@citea{,\penalty\@m\ }{\bf ?}\@warning
       {Citation `\@citeb' on page \thepage \space undefined}}%
    {\setbox\z@\hbox{\global\@tempcntc0\csname b@\@citeb\endcsname\relax}%
     \ifnum\@tempcntc=\z@ \@citeo\@tempcntb\m@ne
       \@citea\def\@citea{,\penalty\@m}
       \hbox{\csname b@\@citeb\endcsname}%
     \else
      \advance\@tempcntb\@ne
      \ifnum\@tempcntb=\@tempcntc
      \else\advance\@tempcntb\m@ne\@citeo
      \@tempcnta\@tempcntc\@tempcntb\@tempcntc\fi\fi}}\@citeo}{#1}}

\def\@citeo{\ifnum\@tempcnta>\@tempcntb\else\@citea
  \def\@citea{,\penalty\@m}%
  \ifnum\@tempcnta=\@tempcntb\the\@tempcnta\else
   {\advance\@tempcnta\@ne\ifnum\@tempcnta=\@tempcntb \else
\def\@citea{--}\fi
    \advance\@tempcnta\m@ne\the\@tempcnta\@citea\the\@tempcntb}\fi\fi}

\xspace

\newcommand\rhobar{\ensuremath{\overline \rho}\xspace}
\newcommand\etabar{\ensuremath{\overline \eta}\xspace}

\newcommand\Ndof{\ensuremath{N_{\mathrm{dof}}}\xspace}
\newcommand\DeltaChi{\ensuremath{\Delta\chi^2}\xspace}

\newcommand{\seffsf}[1]{\sin\!^2\theta^{#1}_{{\rm eff}}}

\newcommand{\sinfeff}{\ensuremath{\seffsf{f}}\xspace}
\newcommand{\sinleff}{\ensuremath{\seffsf{\ell}}\xspace}

\newcommand{\mc}{\ensuremath{\overline{m}_c}\xspace}
\newcommand{\mb}{\ensuremath{\overline{m}_b}\xspace}

\newcommand{\tanb}{\ensuremath{\tan\!\beta}\xspace}

\newcommand{\SParam}     	{\ensuremath{0.04\pm 0.11}\xspace}
\newcommand{\TParam}     	{\ensuremath{0.09\pm 0.14}\xspace}
\newcommand{\UParam}     	{\ensuremath{-0.02\pm 0.11}\xspace}

\newcommand{\STParamCor}	{\ensuremath{+0.92}\xspace}
\newcommand{\SUParamCor}	{\ensuremath{-0.68}\xspace}
\newcommand{\TUParamCor}	{\ensuremath{-0.87}\xspace}

\newcommand{\SParamNU}     {\ensuremath{0.04\pm 0.08}\xspace}
\newcommand{\TParamNU}     {\ensuremath{0.08\pm 0.07}\xspace}
\newcommand{\STParamCorNo} {\ensuremath{+0.92}\xspace}

\makeindex
\begin{document}


%
%
\pagenumbering{arabic}
{\small
\color{mygray}
\begin{flushright}
{\sf\em \today} \\
\def\UrlFont{\sf\em}
\url{http://cern.ch/gfitter} 
\end{flushright}
}
\def\UrlFont{\rm}

\vspace{1.3cm}


{\sf\LARGE\bfseries
Update of the global electroweak fit and constraints\\[0.15cm]
on two-Higgs-doublet models
}

\vspace{1.0cm}

{\large \em 
  The Gfitter Group \\[0.2cm]
}
{\large
  J.~Haller$^{a}$, A.~Hoecker$^{b}$, R.~Kogler$^{a}$, 
  K.~M\"onig$^{c}$, T.~Peiffer$^{d}$, J.~Stelzer$^{b}$
}

\vspace{0.2cm}

\begin{details}
  $^{a}$Institut f\"ur Experimentalphysik, Universit\"at Hamburg, Germany\\
  $^{b}$CERN, Geneva, Switzerland \\
  $^{c}$DESY, Hamburg and Zeuthen, Germany \\ 
  $^{d}$II. Physikalisches Institut, Georg-August-Universit\"at G\"ottingen, Germany
\end{details}

\vspace{2.0cm}

\begin{details} {\sf\bfseries Abstract} 
  --- We present an update of the global fit of the Standard Model
electroweak sector to latest experimental results. 
We include new kinematic top quark and 
$W$ boson mass measurements from the LHC, a \sinleff result
from the Tevatron, and a new evaluation of the hadronic contribution
to $\alpha(M_Z^2)$. We present tests of the internal consistency of 
the electroweak Standard Model and updated numerical predictions of key observables.
The electroweak data combined with measurements of the Higgs boson coupling
strengths and flavour physics observables are  used  to constrain
parameters of two-Higgs-doublet models.  
\end{details}

\thispagestyle{empty}

\newpage
\tableofcontents
\newpage

%
%
\section{Introduction}
\label{sec:intro}

\begin{figure}[p]
\begin{center}
\includegraphics[width=\defaultSingleFigureScale\textwidth]{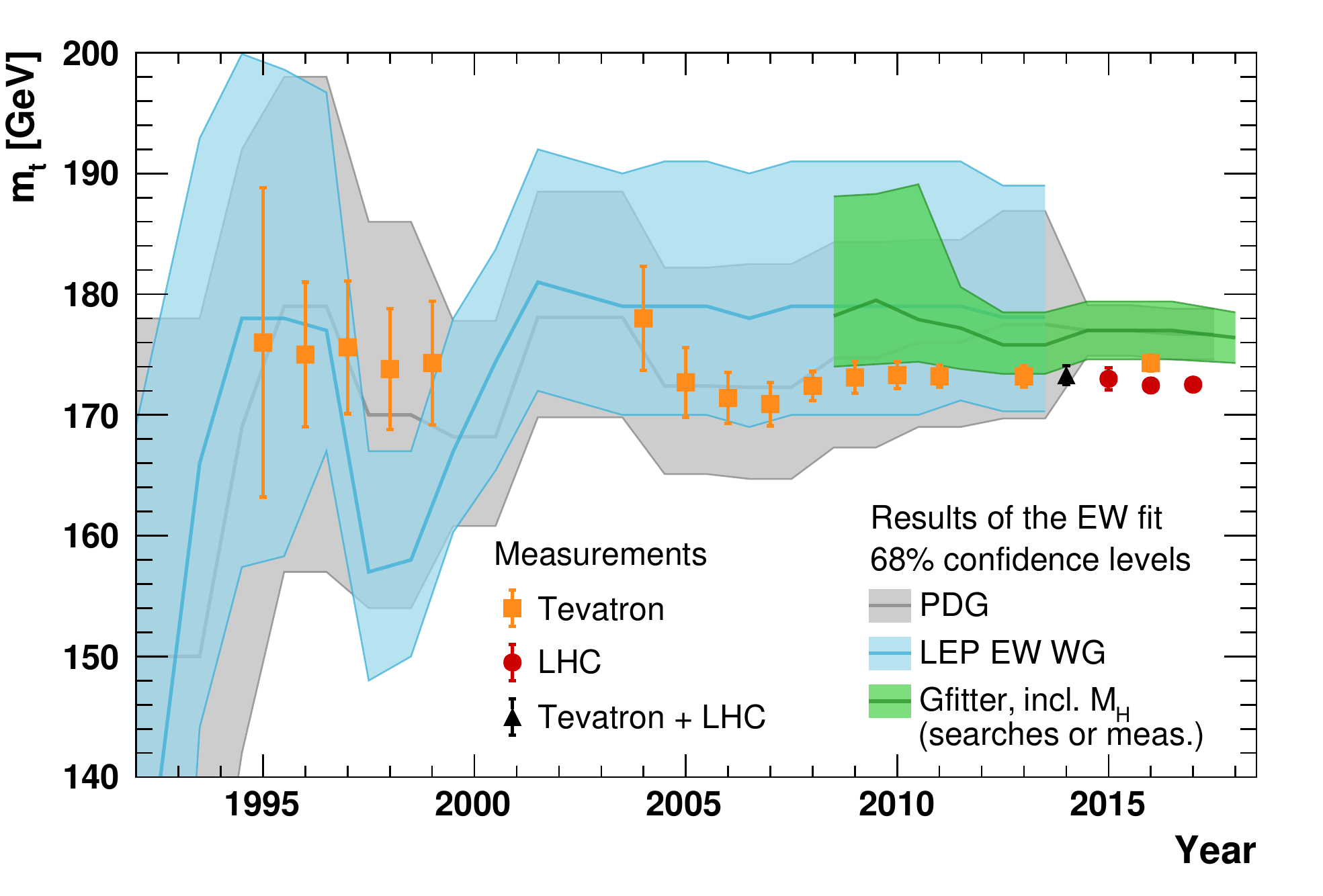}
\end{center}
\vspace{-0.3cm}
\caption[]{Prediction of the top quark mass versus year as obtained by various analysis
  groups using electroweak precision data (grey~\cite{pdgweb}, light blue~\cite{lepppe}, 
  green~\cite{GfitterWeb}). 
  The bands indicate the 68\% confidence level. 
  The direct $m_t$ measurements after the top quark discovery are displayed by the data points 
  (orange~\cite{Abachi:1995iq, Abe:1995hr, tevewwg, TevatronElectroweakWorkingGroup:2016lid, Abazov:2017ktz}, red~\cite{Aad:2015nba, Khachatryan:2015hba, ATLAS:2017lqh}, black~\cite{ATLAS:2014wva}). }
\label{fig:mt_history}
\end{figure}

\begin{figure}[p]
\begin{center}
\includegraphics[width=\defaultSingleFigureScale\textwidth]{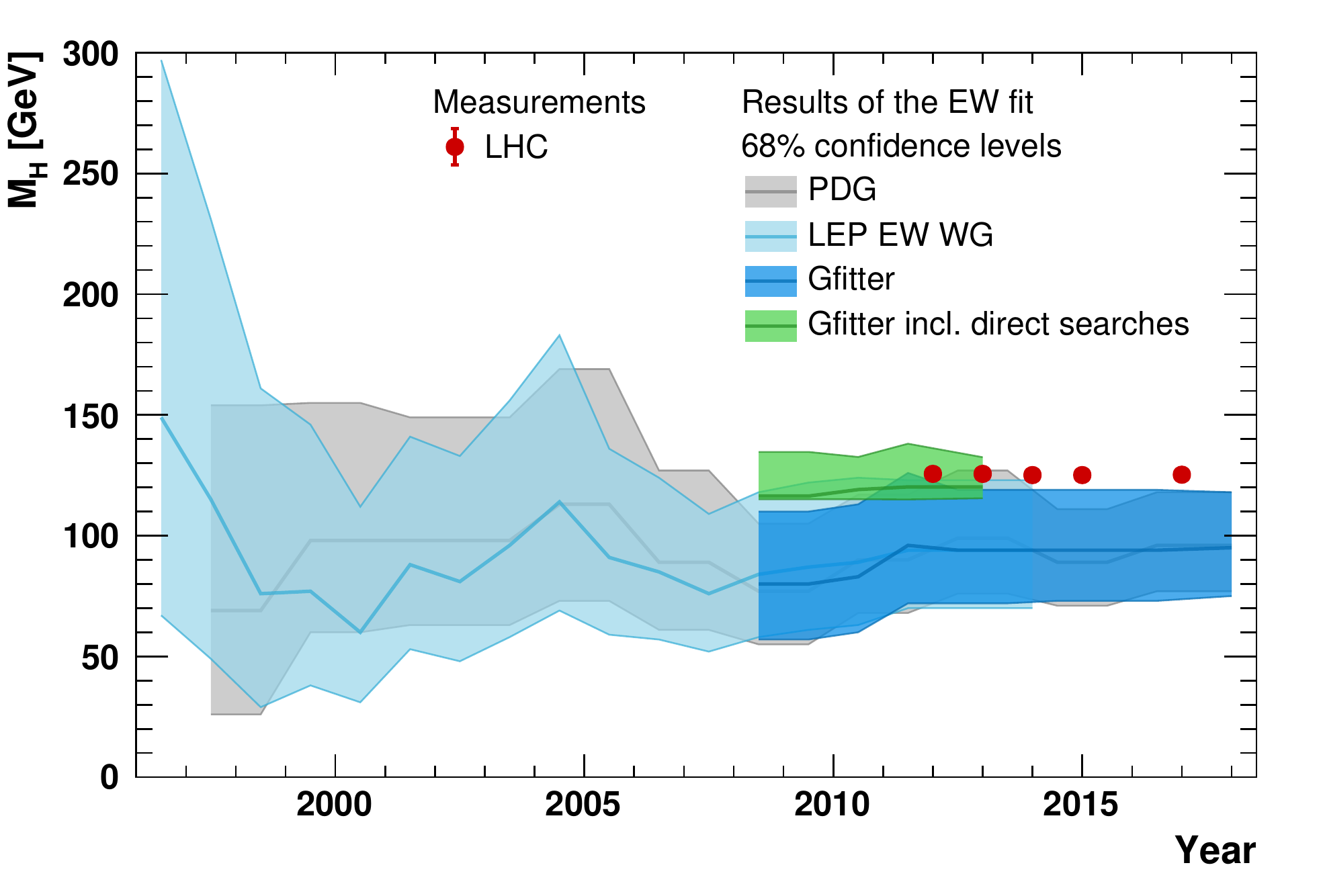}
\end{center}
\vspace{-0.3cm}
\caption[]{Prediction of the Higgs boson mass versus year as obtained by various analysis
  groups using electroweak precision data (grey~\cite{pdgweb}, light blue~\cite{lepppe}, dark
  blue~\cite{GfitterWeb}) and including direct search results (green~\cite{GfitterWeb}).  
  The bands indicate the 68\% confidence level.
  The direct $M_H$ measurements after the Higgs boson discovery are displayed by the red 
  data points~\cite{Aad:2012tfa,Chatrchyan:2012xdj,Aad:2015zhl, ATLAS-CONF-2017-046, Sirunyan:2017exp}. }
\label{fig:mh_history}
\end{figure}

Since the 1990'ies, electroweak precision data 
from LEP and SLD~\cite{ALEPH:2005ema} were used together with accurate
Standard Model (SM) calculations to predict parameters of the theory. 
A first impressive confirmation of the predictive power of global fits 
in high-energy physics (HEP) was the discovery of the top quark 
at the Tevatron~\cite{Abachi:1995iq, Abe:1995hr} in 1995, with a mass 
in agreement with the predictions from global fits. 
Knowledge of the top quark mass ($m_t$) made it possible 
to constrain the mass of the Higgs boson ($M_H$). Increasing experimental 
and theoretical precision and the inclusion of constraints from direct 
Higgs boson searches from LEP and Tevatron narrowed the allowed mass 
range over time~\cite{lepppe,pdgweb,GfitterWeb,HEPFitterWeb,Flacher:2008zq}. 
The discovery of the Higgs boson at the 
Large Hadron Collider (LHC)~\cite{Aad:2012tfa,Chatrchyan:2012xdj} with a mass around 
$125\:$GeV impressively confirmed the SM at the quantum level. 
The historical development of the constraints is
illustrated in Figs.~\ref{fig:mt_history} and \ref{fig:mh_history}, 
where the predictions of, respectively, $m_t$ and $M_{H}$ as derived from various global 
fits and direct measurements~\cite{Abachi:1995iq, Abe:1995hr, tevewwg, 
ATLAS:2014wva, Aad:2015nba, Khachatryan:2015hba, TevatronElectroweakWorkingGroup:2016lid, Abazov:2017ktz, Aad:2012tfa,Chatrchyan:2012xdj,Aad:2015zhl, ATLAS-CONF-2017-046, Sirunyan:2017exp, ATLAS:2017lqh} 
are shown versus time.

With the measurement of $M_{H}$ the electroweak sector of
the SM is overconstrained and the strength of global fits can be
exploited to predict key observables such as the $W$ boson mass and
the effective electroweak mixing angle, with a precision exceeding that of the
direct measurements~\cite{Baak:2012kk}. Since the last update of our
fit~\cite{Baak:2014ora} improved experimental results have become
available that allow for more accurate tests of the internal
consistency of the SM. Among these are the first
determination of the $W$ boson mass at the LHC
by the ATLAS collaboration~\cite{Aaboud:2017svj},
new combined results of the top quark mass by the 
LHC experiments~\cite{Khachatryan:2015hba, ATLAS:2017lqh},
a new combination of measurements of the
effective leptonic electroweak mixing angles from the Tevatron
experiments~\cite{Aaltonen:2018dxj}, a 
Higgs boson mass combination released by the ATLAS and 
CMS collaborations~\cite{Aad:2015zhl}, 
and an updated value
of the hadronic contribution to the running of the electromagnetic
coupling strength at the $Z$ boson mass~\cite{Davier:2017zfy}.  
In the first part of this paper we present an update of the electroweak 
fit including these new experimental
results and up-to-date theoretical predictions.

While the Higgs boson  measurements  so far agree 
with a minimal scalar  sector as implemented in the SM, the question 
remains whether a more complex scalar sector may be realised in
nature, possibly featuring a variety of Higgs boson states.
Two-Higgs-doublet models (2HDM)~\cite{Haber:1978jt} are a popular SM
extension in which an additional $SU(2)_L\times
U(1)_Y$ scalar doublet field with hypercharge $Y=1$ is added to the SM
leading to the existence of five physical Higgs boson states, $h$, $H$,
$A$, $H^+$, and $H^-$, where the neutral $h$ may be identified
with the discovered 125$\:$GeV Higgs boson as is assumed in
this paper. The scalar $H$ boson has CP-even quantum number, $A$ is
a CP-odd pseudo-scalar, and $H^+$ and $H^-$ carry opposite electric 
charge but have identical mass. 
No experimental hint for additional scalar states has been observed 
so far in direct searches~\cite{Abbiendi:2013hk, 
Abazov:2008rn, Abazov:2009aa, Aaltonen:2009ke, 
Aad:2013hla,Aad:2014kga,Aad:2015typ,Aad:2015nfa, Aaboud:2016cre,
Aaboud:2017rel,Aaboud:2017cxo,Aaboud:2017sjh,Aaboud:2017hnm, 
Chatrchyan:2012vca,Khachatryan:2015qxa, Khachatryan:2015uua,
Khachatryan:2016are}.
In this situation global 2HDM fits, exploiting observables  sensitive
to these additional Higgs boson states via quantum corrections,
can be used to constrain the allowed mass ranges and  2HDM mixing
parameters. In the second part of this article such constraints are 
derived from a global fit using a combination of electroweak precision
data, flavour physics observables, the anomalous magnetic moment of the muon, and
measurements of the Higgs boson coupling strength to SM particles.  


\section{Update of the global electroweak fit}
\label{sec:ewfit}

The updated global electroweak fit presented in this section uses the Gfitter
framework. For a detailed discussion of the experimental
data, the implementation of the theoretical predictions, and the
statistical procedure employed by Gfitter we refer the reader to our previous 
publications~\cite{Flacher:2008zq,Baak:2011ze,Baak:2012kk,Baak:2014ora}. 
A detailed list of all the observables, their values and uncertainties used 
in the fit, is given in the first two columns of Table~\ref{tab:results}.
The description below discusses recent changes in the input quantities 
and calculations.

\subsection{Input measurements and theoretical predictions}
\label{sec:impr}

The electroweak precision data measured at the $Z$ pole and their
correlations~\cite{ALEPH:2005ema} as well as the width of the $W$
boson have not changed since
our last analysis~\cite{Baak:2014ora}. The update to the
most recent world average values for the running $c$ and $b$ quark
masses~\cite{Olive:2016xmw} has negligible impact on the fit
result. This is also the case for the Run-1 LHC average
of the Higgs boson mass, $M_H=125.09\pm0.21\pm0.11\:\gev$~\cite{Aad:2015zhl},
which we use now instead of a simple weighted average.\footnote{The Run-1 result on $M_H$ 
was confirmed by ATLAS and CMS measurements at 
$\sqrt{s} = 13\:\gev$~\cite{ATLAS-CONF-2017-046,Sirunyan:2017exp}.}

New results are  available for several observables with
high sensitivity and potentially significant impact on the fit. 
We include new measurements of the $W$ boson and top quark 
masses as described in the following sections. For the
first time we include as a separate fit input (assuming no correlation with other measurements) the latest combination
of measurements of the effective leptonic electroweak mixing angle
from the Tevatron experiments\footnote{The \sinleff measurements of ATLAS 
($\sinleff=0.2308\pm0.0012$~\cite{Aad:2015uau}) and CMS ($\sinleff=0.23101\pm0.00052$~\cite{CMS:2017zzj}) are not 
included in the fit because of their presently insufficient precision and unknown correlations.},
$\sinleff=0.23148\pm0.00033$~\cite{Aaltonen:2018dxj},
and we use an updated value for the five quark flavour hadronic contribution to the
running of the electromagnetic coupling strength at $M_Z$,
$\dahadZf=(2760\pm9)\cdot10^{-5}$~\cite{Davier:2017zfy}.

\subsubsection*{{\em W} boson mass} 

The ATLAS collaboration has recently released the first LHC measurement
of the mass of the $W$ boson~\cite{Aaboud:2017svj}. Analysing their 
$7\:\tev$ dataset ATLAS measures $M_W = 80\,370 \pm 7_{\rm stat} \pm 11_{\rm exp\;syst} \pm
14_{\rm model}\:\mev$. We include this result in the
fit by combining it with the Tevatron 
($M_W = 80\,387\pm 16\:\mev$~\cite{TevatronElectroweakWorkingGroup:2012gb}) 
and LEP combinations 
($M_W = 80\,376 \pm 25_{\rm stat} \pm 22_{\rm syst}\:\mev$~\cite{LEPEWWG}) as follows.

Using information from Ref.~\cite{TevatronElectroweakWorkingGroup:2012gb} 
we estimate the composition of individual statistical, experimental systematic 
and modelling uncertainties in the combined Tevatron 
result by $\pm8_{\rm stat} \pm 8_{\rm exp\;syst} \pm 12_{\rm model}\:\mev$. 
All statistical and experimental systematic uncertainties are assumed to be 
uncorrelated among the three input results (ATLAS, Tevatron, LEP) as is the 
modelling uncertainty from LEP.
The impact of the unknown correlation 
among the modelling uncertainties affecting the ATLAS and Tevatron measurements 
has been studied by varying its value between zero and one. For a large range of
correlations we observe a stable
average of $M_W = 80\,379 \pm 13\:\mev$, which we use in the fit.\footnote{A
      central value of $80\,379\:\mev$ is obtained for all possible
      values of the model correlation, except for 
      coefficients exceeding 0.9 for which a value of $80\,380\:\mev$ is
      found. A combined uncertainty of $13\:\mev$ is obtained for 
      correlations  between 0.4 and 0.9, while smaller and larger correlation
      values yield $12\:\mev$ and $14\:\mev$, respectively. These values have
      been consistently calculated using the Best Linear Unbiased
      Estimate (BLUE)~\cite{Valassi:2003mu} and the least-squares
      averaging implemented in Gfitter~\cite{Flacher:2008zq}.} 


\subsubsection*{Top quark mass}

For lack of a recent \mt world average, we attempt here for the purpose
of the fit a conservative combination 
of the most precise kinematic \mt measurements obtained at the LHC.
We combine the $\mt$ averages from ATLAS
($172.51 \pm 0.27_\mathrm{stat} \pm 0.42_\mathrm{syst}\:\gev$)~\cite{ATLAS:2017lqh} 
and CMS
($172.47 \pm 0.13_\mathrm{stat} \pm 0.47_\mathrm{syst}\:\gev$)~\cite{Khachatryan:2015hba}, 
which are based on 7 and 8\:\tev data. These averages include  results from the 
dilepton~\cite{Aaboud:2016igd, 
Chatrchyan:2011nb, Chatrchyan:2012ea}, 
lepton+jets~\cite{Aad:2015nba, Chatrchyan:2012cz} and 
fully hadronic~\cite{Chatrchyan:2013xza} channels.
Assuming the overlapping fraction of the systematic uncertainties 
to be fully correlated (which corresponds to a correlation coefficient 
of 72\% between the two measurements)
we obtain the combined value
$\mt = 172.47 \pm 0.46\:\gev$ ($p$-value of $0.84$), which we use as input in the fit.

The latest average from the D0 collaboration $\mt = 174.95 \pm
0.40_\mathrm{stat} \pm 0.64_\mathrm{syst}\:\gev$~\cite{Abazov:2017ktz}
is barely compatible with the aforementioned average of the LHC
measurements.  A combination of the D0 average with the LHC average
would result in $p$-values between $5\cdot10^{-3}$ and $3\cdot10^{-5}$,
depending on the assumed correlation between the systematic uncertainties.
The result from the CDF collaboration,  $\mt = 173.16
\pm 0.57_\mathrm{stat} \pm
0.74_\mathrm{syst}\,\gev$~\cite{CDF:mtopcombo}, agrees with
the LHC average, with $p$-values between $0.40$ and $0.51$ depending 
on the correlation.

As in our previous work~\cite{Baak:2014ora} we assign an additional theoretical
uncertainty of $0.5\:\gev$ to the value of $\mt$ from hadron collider
measurements due to the ambiguity in the kinematic top quark mass
definition~\cite{Hoang:2008yj,Hoang:2008xm,Buckley:2011ms,Moch:2014tta,
  top13mangano}, the colour structure of the fragmentation
process~\cite{Skands:2007zg,Wicke:2008iz}, and the perturbative
relation between pole and \MSbar mass currently known to three-loop
order~\cite{Chetyrkin:1999qi, Melnikov:2000qh, Hoang:2000yr}.

\subsubsection*{Theoretical calculations}

The theoretical higher-order calculations used in Gfitter have not changed since 
our last update~\cite{Baak:2014ora}, except for new bosonic two-loop corrections 
to the $Zb\overline{b}$ vertex~\cite{Dubovyk:2016aqv}.

For the effective weak mixing angle \sinfeff we use the parametrisations provided 
in~\cite{Awramik:2004ge, Awramik:2006uz, Dubovyk:2016aqv}, which include full 
two-loop electroweak~\cite{Awramik:2004ge, Awramik:2006uz} and partial three-loop and 
four-loop QCD corrections~\cite{Avdeev:1994db, Chetyrkin:1995ix, Chetyrkin:1995js, 
vanderBij:2000cg, Faisst:2003px, Schroder:2005db, Chetyrkin:2006bj, Boughezal:2006xk}. 
For bottom quarks, the calculations from Refs.~\cite{Dubovyk:2016aqv, Awramik:2008gi} are used. 
The new bosonic two-loop corrections are numerically small. They shift the prediction 
of the forward-backward asymmetry for $b$ quarks $A_{\rm FB}^{0,b}$ by $1.3\cdot10^{-5}$, 
which is two orders of magnitude smaller than the experimental uncertainty and thus
does not alter the fit results.  
We use the parametrisation of the full two-loop result~\cite{Awramik:2003rn} 
for predicting the mass of the $W$ boson, where we also include four-loop QCD 
corrections~\cite{Schroder:2005db, Chetyrkin:2006bj, Boughezal:2006xk}.
Full fermionic two-loop corrections for the partial widths and branching ratios 
of the $Z$ boson and the hadronic peak cross section $\sigma^0_{\rm had}$ 
are used~\cite{Freitas:2014hra, Freitas:2013dpa, Freitas:2012sy}.
The dominant contributions from final-state QED and QCD radiation
are included in the calculations~\cite{Chetyrkin:1994js, Baikov:2008jh, Baikov:2012er, 
Kataev:1992dg, Czarnecki:1996ei, Harlander:1997zb}.
The width of the $W$ boson is known up to one electroweak 
loop order, where we use the parametrisation given in Ref.~\cite{Cho:2011rk}.

The size and treatment of theoretical uncertainties are unchanged with respect 
to our last analysis~\cite{Baak:2014ora}.


\subsection{Results}
\label{sec:sm}

\begin{table}
\setlength{\tabcolsep}{0.0pc}
{\small
\begin{tabular*}{\textwidth}{@{\extracolsep{\fill}}lccccc} 
\hline\noalign{\smallskip}
& & Free &  & \multic{1}{c}{Fit w/o exp. input} & \multic{1}{c}{Fit w/o exp. input}   \\[-0.1cm]
\rs{Parameter} & \rs{Input value} & in fit & \rs{Fit Result} & \multic{1}{c}{in line} & \multic{1}{c}{in line, no theo. unc.} \\
\noalign{\smallskip}\hline\noalign{\smallskip}
$M_{H}$ {\ft [GeV]} &  $125.1\pm0.2$ & yes & $125.1\pm0.2$ & $90^{\,+21}_{\,-18}$ & $89^{\,+20}_{\,-17}$\\
\noalign{\smallskip}\hline\noalign{\smallskip}
$M_{W}$ {\ft [GeV]} &  $80.379\pm0.013$ & -- &  $80.359\pm0.006$ &  $80.354\pm0.007$ &  $80.354\pm0.005$\\
$\Gamma_{W}$ {\ft [GeV]} &  $2.085\pm0.042$ & -- &  $2.091\pm0.001$ &  $2.091\pm0.001$ &  $2.091\pm0.001$\\
\noalign{\smallskip}\hline\noalign{\smallskip}
$M_{Z}$ {\ft [GeV]} &  $91.1875\pm0.0021$ & yes &  $91.1882\pm0.0020$ &  $91.2013\pm0.0095$ &  $91.2017\pm0.0089$\\
$\Gamma_{Z}$ {\ft [GeV]} &  $2.4952\pm0.0023$ & -- &  $2.4947\pm0.0014$ &  $2.4941\pm0.0016$ &  $2.4940\pm0.0016$\\
$\sigma_{\rm had}^{0}$ {\ft [nb]} &  $41.540\pm0.037$ & -- &  $41.484\pm0.015$ &  $41.475\pm0.016$ &  $41.475\pm0.015$\\
$R^{0}_{\l}$ &  $20.767\pm0.025$ & -- &  $20.742\pm0.017$ &  $20.721\pm0.026$ &  $20.719\pm0.025$\\
$A_{\rm FB}^{0,\l}$ &  $0.0171\pm0.0010$ & -- &  $0.01620\pm0.0001$ &  $0.01619\pm0.0001$ &  $0.01619\pm0.0001$\\
 $A_\ell$ $^{(\star)}$  & $0.1499\pm0.0018$ & --  & $0.1470\pm0.0005$ & $0.1470\pm0.0005$ & $0.1469\pm0.0003$\\
$\sinleff(Q_{\rm FB})$ &  $0.2324\pm0.0012$ & -- &  $0.23153\pm0.00006$ &  $0.23153\pm0.00006$ &  $0.23153\pm0.00004$\\
$\sinleff(\rm Tevt.)$ &  $0.23148\pm0.00033$ & -- &  $0.23153\pm0.00006$ &  $0.23153\pm0.00006$ &  $0.23153\pm0.00004$\\
$A_{c}$ &  $0.670\pm0.027$ & -- &  $0.6679\pm0.00021$ &  $0.6679\pm0.00021$ &  $0.6679\pm0.00014$\\
$A_{b}$ &  $0.923\pm0.020$ & -- &  $0.93475\pm0.00004$ &  $0.93475\pm0.00004$ &  $0.93475\pm0.00002$\\
$A_{\rm FB}^{0,c}$ &  $0.0707\pm0.0035$ & -- &  $0.0736\pm0.0003$ &  $0.0736\pm0.0003$ &  $0.0736\pm0.0002$\\
$A_{\rm FB}^{0,b}$ &  $0.0992\pm0.0016$ & -- &  $0.1030\pm0.0003$ &  $0.1032\pm0.0003$ &  $0.1031\pm0.0002$\\
$R^{0}_{c}$ &  $0.1721\pm0.0030$ & -- &  $0.17224\pm0.00008$ &  $0.17224\pm0.00008$ &  $0.17224\pm0.00006$\\
$R^{0}_{b}$ &  $0.21629\pm0.00066$ & -- &  $0.21582\pm0.00011$ &  $0.21581\pm0.00011$ &  $0.21581\pm0.00004$\\
\noalign{\smallskip}\hline\noalign{\smallskip}
$\mc$ {\ft [GeV]} &  $1.27^{\,+0.07}_{\,-0.11}$ & yes &  $1.27^{\,+0.07}_{\,-0.11}$ & --  & -- \\
$\mb$ {\ft [GeV]} &  $4.20^{\,+0.17}_{\,-0.07}$ & yes &  $4.20^{\,+0.17}_{\,-0.07}$ & --  & -- \\
$m_{t}$ {\ft [GeV]}$^{(\bigtriangledown)}$ &  $172.47\pm0.68$ & yes &  $172.83\pm0.65$ &  $176.4\pm2.1$ &  $176.4\pm2.0$\\
$\dalphaHadMZ$ $^{(\dag\bigtriangleup)}$ &  $2760\pm   9$ & yes & $2758\pm  9$ & $2716\pm  39$ & $2715\pm  37$\\
$\alpha_{s}(M_{Z}^{2})$ & -- & yes &  $0.1194\pm0.0029$ &  $0.1194\pm0.0029$ &  $0.1194\pm0.0028$\\
\noalign{\smallskip}\hline
\noalign{\smallskip}
\end{tabular*}
{\ft
$^{(\star)}$Average of LEP ($A_\ell=0.1465\pm0.0033$) and SLD ($A_\ell=0.1513\pm0.0021$) measurements, used as two measurements in the fit.
The fit without the LEP (SLD) measurement gives $A_\ell=$ $0.1470\pm0.0005$ ($A_\ell=$ $0.1467\pm0.0005$).
$^{(\bigtriangledown)}$Combination of experimental (0.46~GeV) and theory uncertainty (0.5~GeV).$^{(\dag)}$In units of $10^{-5}$.
$^{(\bigtriangleup)}$Rescaled due to $\alpha_s$ dependency.
}}
\vspace{0.4cm}
\caption{Input values and fit results for the observables used in the global electroweak fit.
The first and second columns list respectively the observables/parameters used in the fit,
and their experimental values or phenomenological estimates (see text for references).
The third column indicates whether a parameter is floating in the fit. The fourth column
gives the results of the fit including all experimental data. In the fifth column, the
fit results are given without using the corresponding experimental or phenomenological
estimate in the given row (indirect determination). The last column shows for illustration
the result using the
same fit setup as in the fifth column, but ignoring all theoretical uncertainties.
\label{tab:results}
}
\end{table}

The fit uses as input observables the quantities and values given in the left rows 
of Table~\ref{tab:results}. The fit parameters are $M_H$, $M_Z$, $m_c$, $m_b$,
\mt, $\dalphaHadMZ$, \as, as well as ten theoretical uncertainty ({\em nuisance}) 
parameters constrained by Gaussian functions (see Ref.~\cite{Baak:2014ora} 
for more details). 

The fit results in a minimum $\chi^2$ value of $18.6$ for $15$ degrees
of freedom, corresponding to a $p$-value of $0.23$. The
results of the full fit for each observable are
given in the fourth column of Table~\ref{tab:results}, together with
the uncertainties estimated from their $\Delta \chi^2 = 1$
profiles. The fifth column in Table~\ref{tab:results} gives the
results obtained without using the experimental measurement
corresponding to that row in the fit ({\it indirect determination} 
of the observable). 
The last column in Table~\ref{tab:results} corresponds to
the fits of the previous column but ignoring all theoretical
uncertainties~\cite{Baak:2014ora}.
\begin{figure}
\begin{center}
\includegraphics[width=0.48\textwidth]{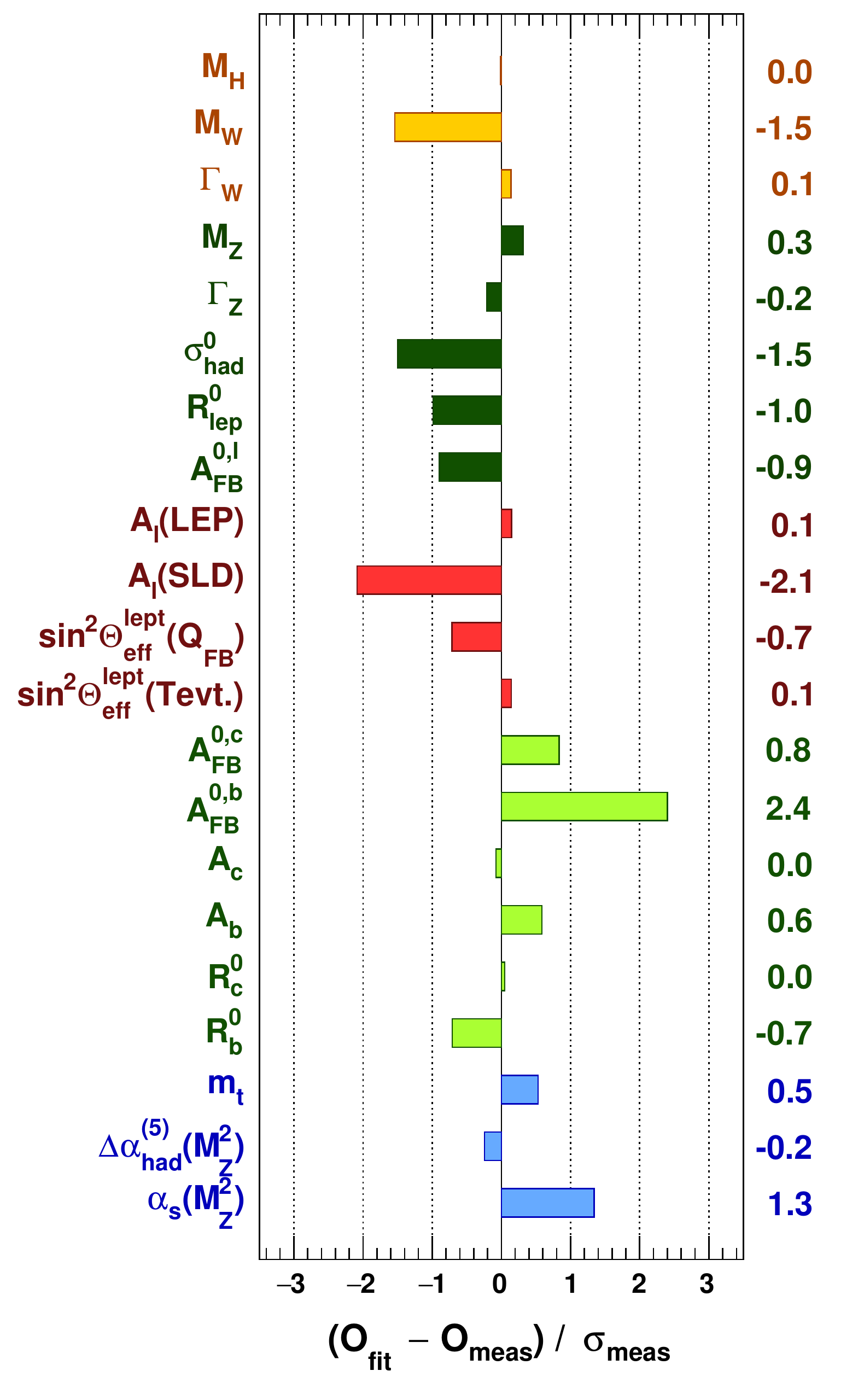}\hspace{0.3cm}
\includegraphics[width=0.46\textwidth]{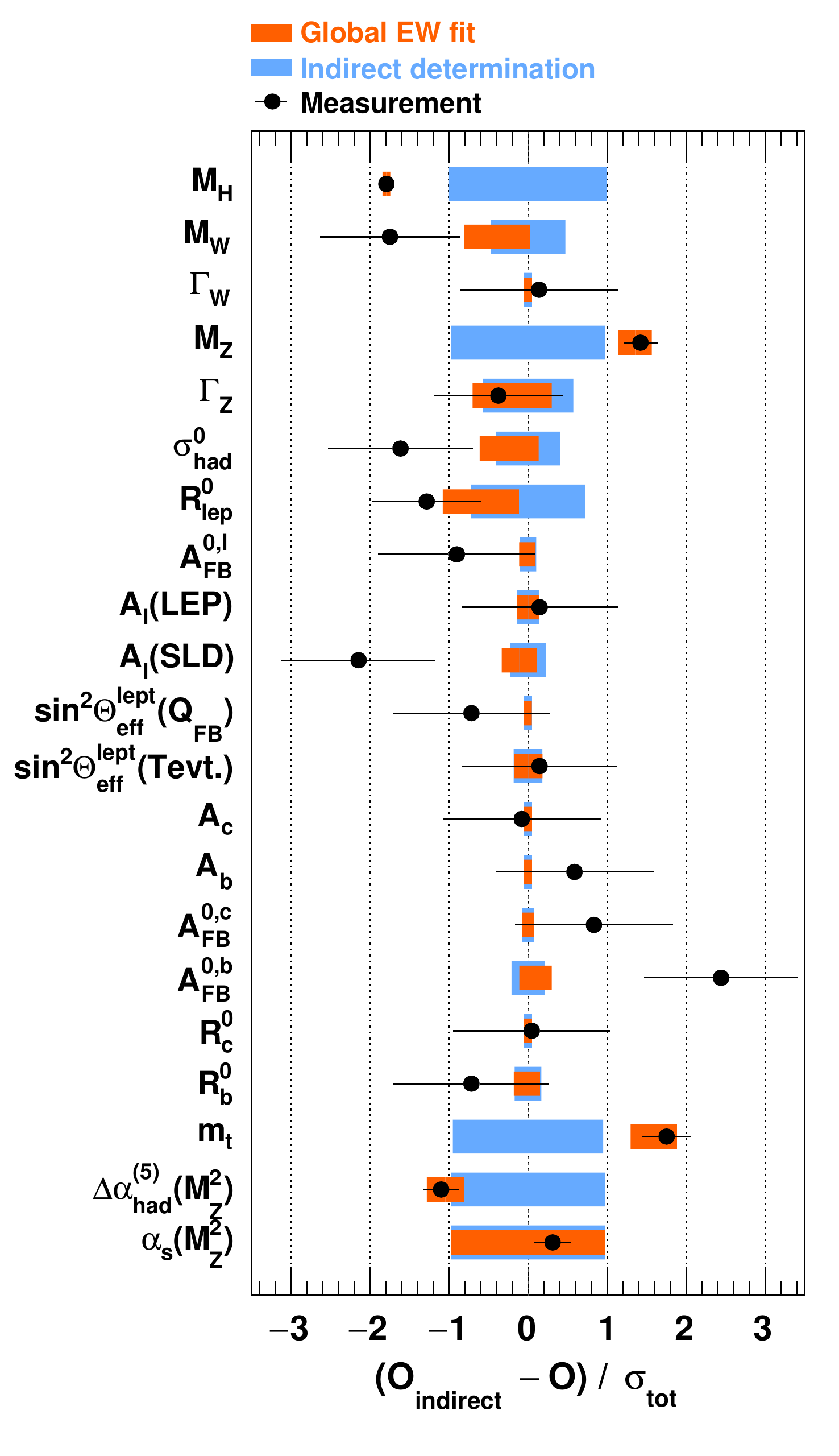}
\end{center}
\vspace{-0.1cm}
\caption[]{Left: comparison of the fit results with the input
  measurements in units of the experimental uncertainties. 
  Right: comparison of the fit results and the input measurements 
  with the indirect determinations in
  units of the total uncertainties. Analog results for the indirect determinations illustrate the impact of their uncertainties on the total uncertainties.
The indirect determination of an
  observable corresponds to a fit without using the constraint from 
  the corresponding input measurement.}
\label{fig:pulls2}
\end{figure}

The left-hand panel of Fig.~\ref{fig:pulls2} displays the  pulls each
given by the difference of the global fit result of an observable (fourth column of 
Table~\ref{tab:results}) and the corresponding input measurement
(second column of Table~\ref{tab:results}) in units of the measurement
uncertainty. The right-hand panel of Fig.~\ref{fig:pulls2} shows 
the difference between the  global fit result (fourth column of 
Table~\ref{tab:results}) as well as the input measurements 
(first column of Table~\ref{tab:results}) with the
indirect determination (fifth column of Table~\ref{tab:results})
for each observable in units of the total uncertainty obtained by adding
in quadrature the uncertainties of the indirect determination and the 
input measurement. 
The analog result using the value of the indirect
determination, trivially centered around zero, are shown to illustrate the impact of its
uncertainty on the total uncertainty.
As in our previous fits, a  tension is observed in the
leptonic and hadronic asymmetry observables, which is largest in
the forward-backward asymmetry of the $b$ quarks, $A^{0,b}_{\rm FB}$. The
impact of the new Tevatron $\sinleff$ measurement on the fit  
result is  small due to   yet insufficient precision.

\begin{figure}
\begin{center}
\includegraphics[width=0.60\textwidth]{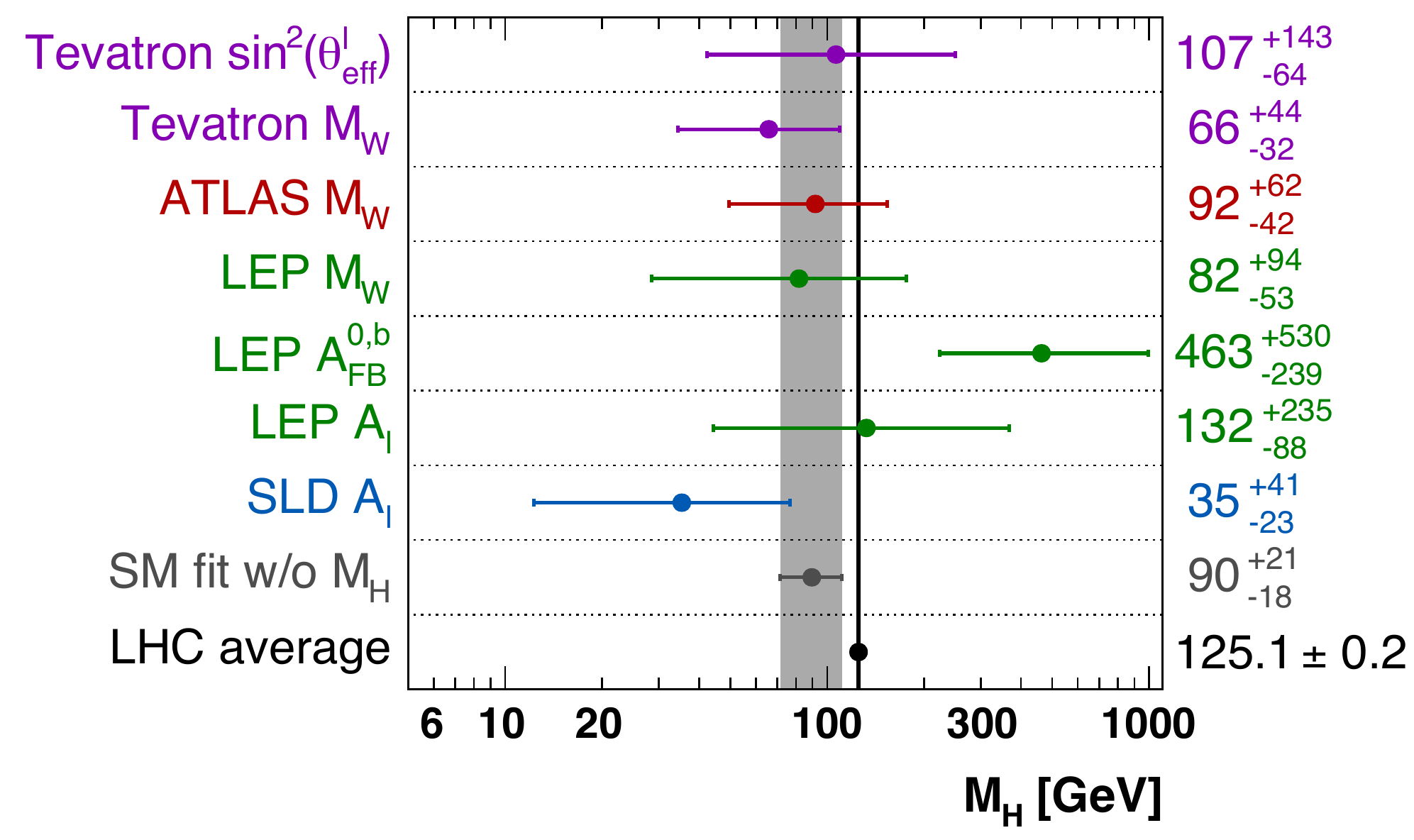}
\end{center}
\vspace{-0.5cm}
\caption[]{Comparison of the constraints on $\MH$ obtained indirectly
  from individual observables with the fit result and the direct
  LHC measurement. For the indirect determinations among the four
  observables providing the strongest $M_H$ constraints (namely
  \sinleff, $M_W$, $A^{0,b}_{\rm FB}$ and $A_\ell$) only the one indicated in
  a given row of the plot is included in the fit. The results shown are not fully independent.  }
\label{fig:mainobs}
\end{figure}
Figure~\ref{fig:mainobs} displays the indirect determination of the Higgs boson
mass from fits in which among the four observables providing
the strongest $M_H$ constraints (namely \sinleff, $M_W$, $A^{0,b}_{\rm FB}$ and $A_\ell$) only the one indicated in a given row of the plot
is included. The results are compared to the direct $M_H$ measurement 
as well as to the result of a fit including all data except the direct $M_H$
measurement. This latter fit gives the indirect determination
\begin{equation}
       M_H = 90^{\,+21}_{\,-18}\:\gev\,,
\label{eq:mh}
\end{equation}
which is in agreement with the direct measurement within $1.7$ 
standard deviations. The value is lower by 
$3\:\gev$ than in our previous result ($93^{\,+25}_{\,-21}\:\gev$)~\cite{Baak:2014ora} 
due to the lower value of $\mt$ used here. The reduced uncertainty of 
$^{\,+21}_{\,-18}\:\gev$ compared to $^{\,+25}_{\,-21}\:\gev$ previously, 
is due to the smaller uncertainty in $m_t$. 
When assuming perfect knowledge of $m_t$, $\dalphaHadMZ$ and $\as(M_{Z}^{2})$, 
the uncertainty is reduced by $\,^{\,+4.5}_{\,-3.5}$, $\,^{\,+5}_{\,-4}$ and 
$\pm2\:\gev$, respectively.
The predictions of $M_H$ using $A_\ell$, $A^{0,b}_{\rm FB}$ and $M_W$ 
(LEP and Tevatron) concur with earlier findings~\cite{Flacher:2008zq}. 
The predictions derived from the ATLAS $M_W$ and  
Tevatron \sinleff measurements are in agreement with the direct 
$M_H$ measurement.

\begin{figure}
\begin{center}
\includegraphics[width=\defaultSingleFigureScale\textwidth]{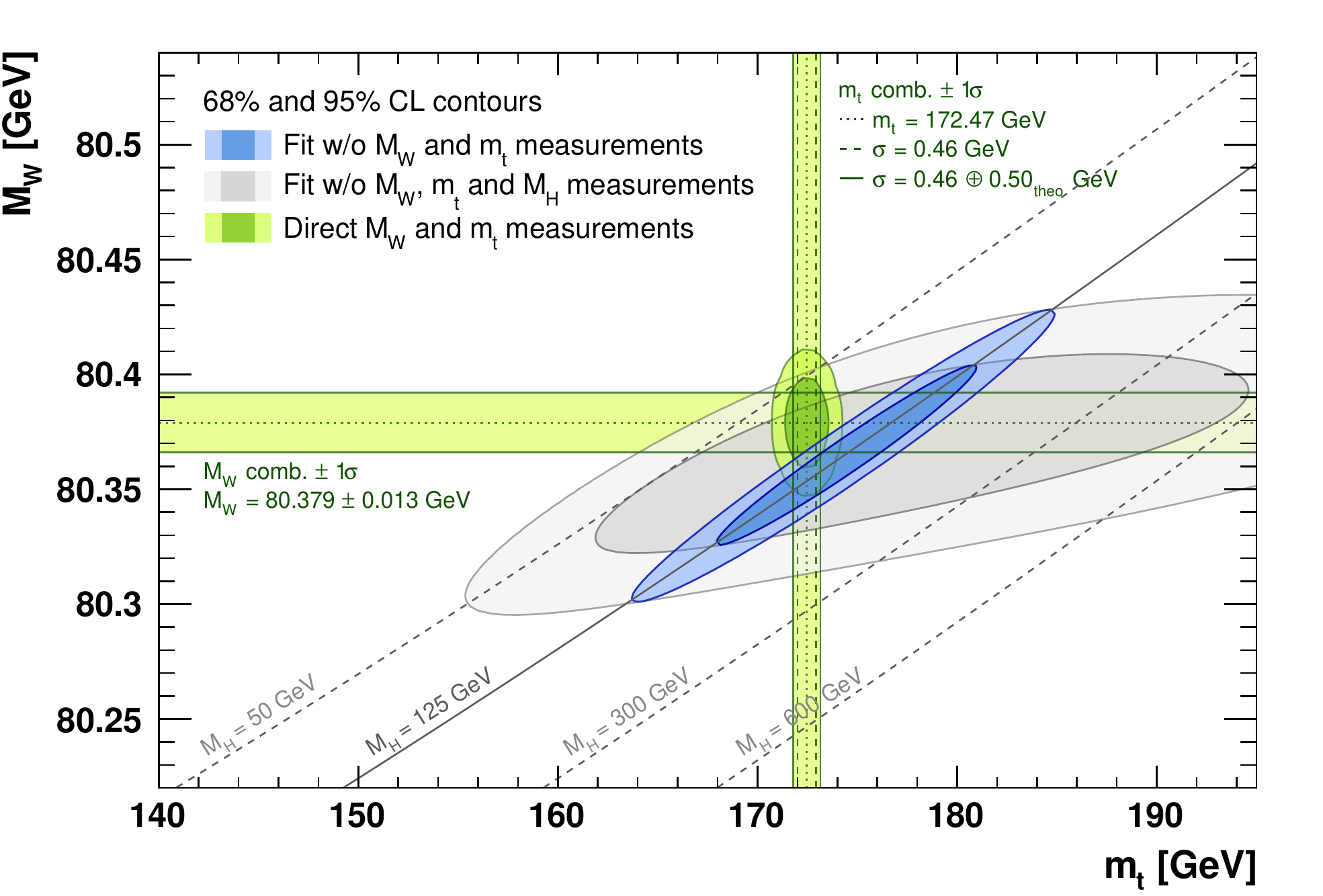}
\end{center}
\vspace{-0.3cm}
\caption[]{Contours at 68\% and 95\% CL obtained from scans of $M_W$
  versus $m_t$ for the fit including (blue) and excluding the $M_H$
  measurement (grey), as compared to the direct measurements (green vertical
  and horizontal $1\sigma$ bands, and two-dimensional $1\sigma$ and $2\sigma$ ellipses). 
  The direct measurements of $M_W$ and $m_t$ are excluded from the fits.  }
\label{fig:mwmt}
\end{figure}

An important consistency test of the SM is the simultaneous indirect
determination of \mt and $M_W$. A scan of the confidence level (CL)
profile of $M_W$ versus $m_t$ is shown in Fig.~\ref{fig:mwmt} for the
scenarios where the direct $M_H$ measurement is included in the fit
(blue) or not (grey). Both contours agree with the direct measurements
(green bands and ellipse for two degrees of freedom).

\begin{figure}
\begin{center}
\includegraphics[width=\defaultDoubleFigureScale\textwidth]{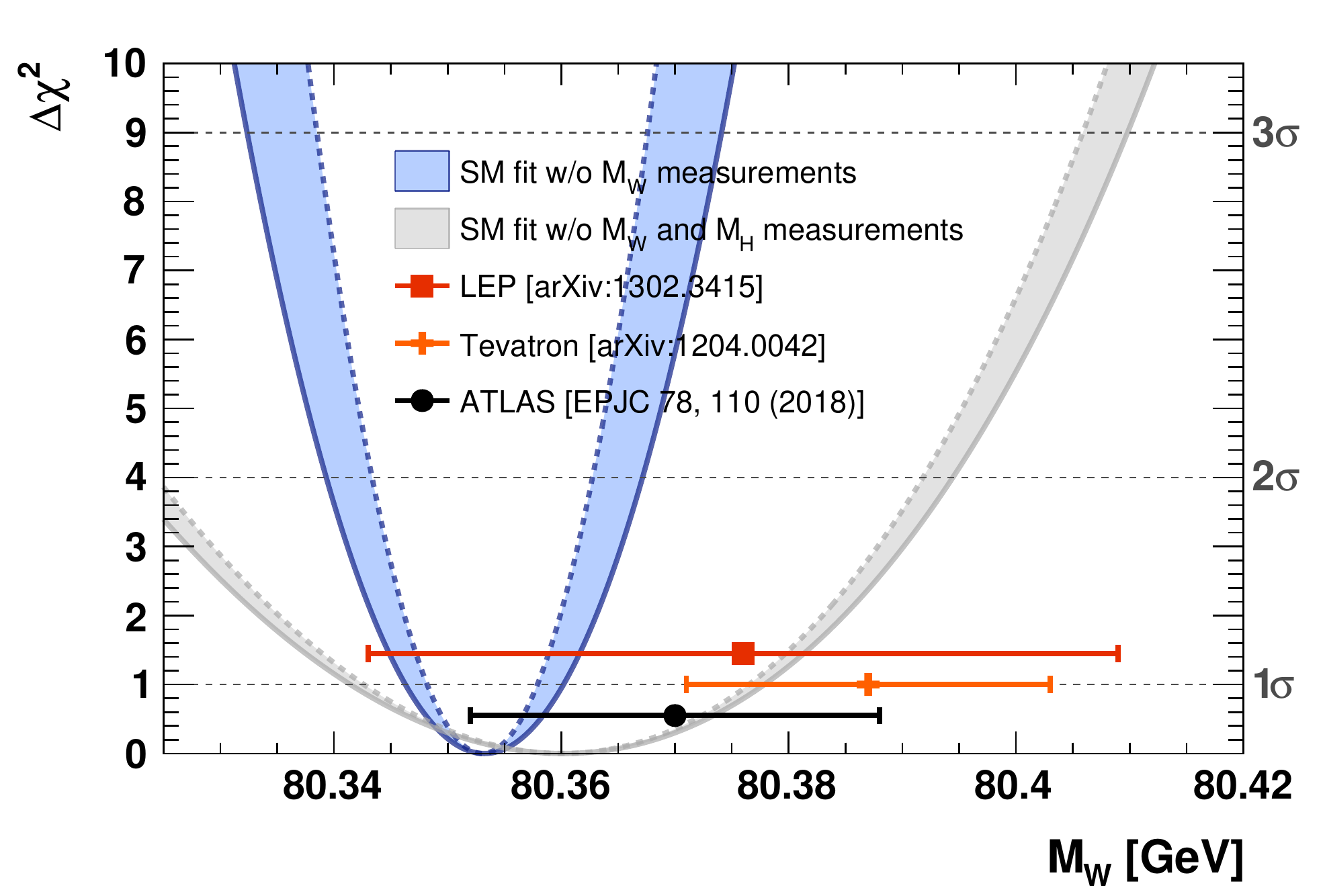}
\includegraphics[width=\defaultDoubleFigureScale\textwidth]{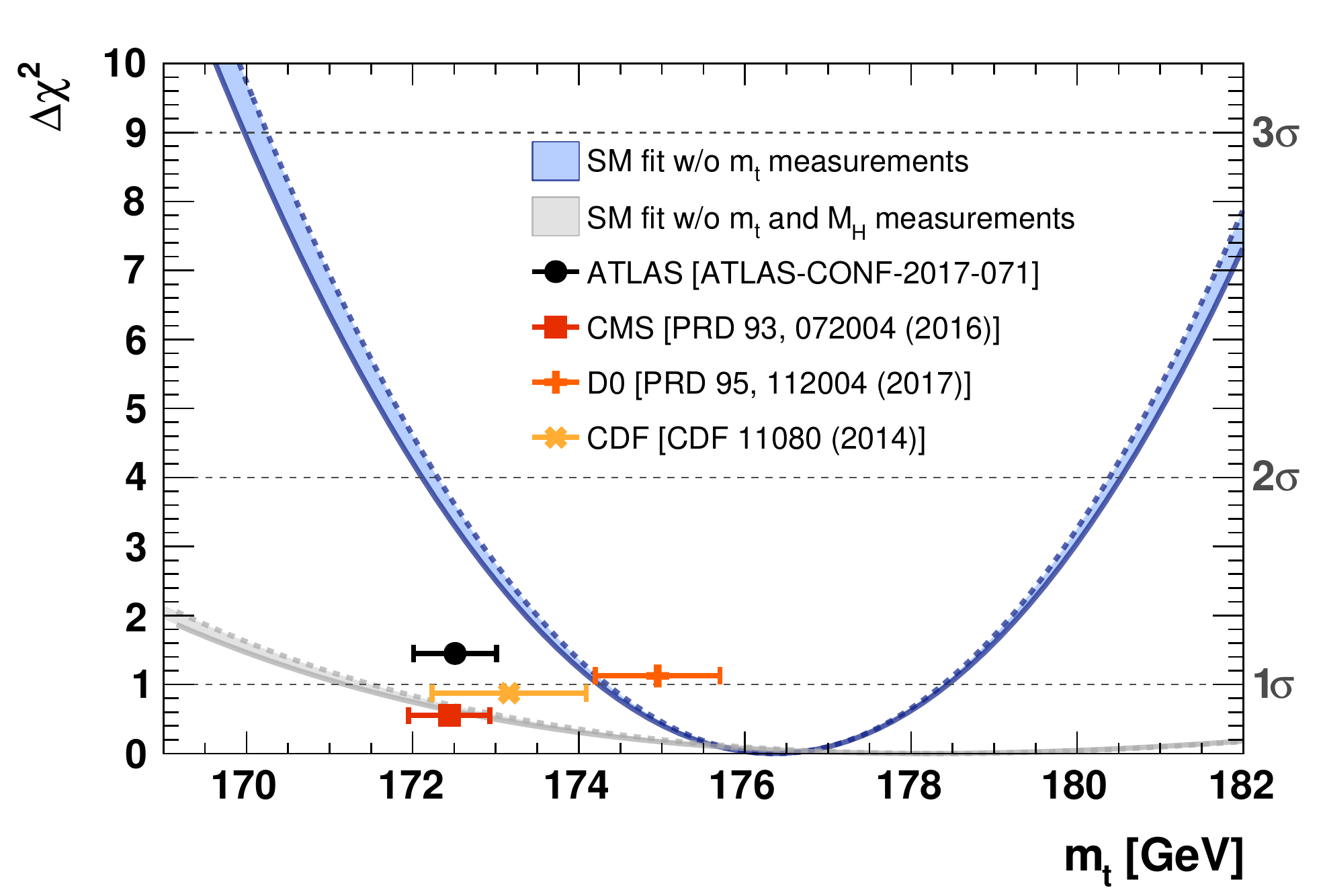}
\includegraphics[width=\defaultDoubleFigureScale\textwidth]{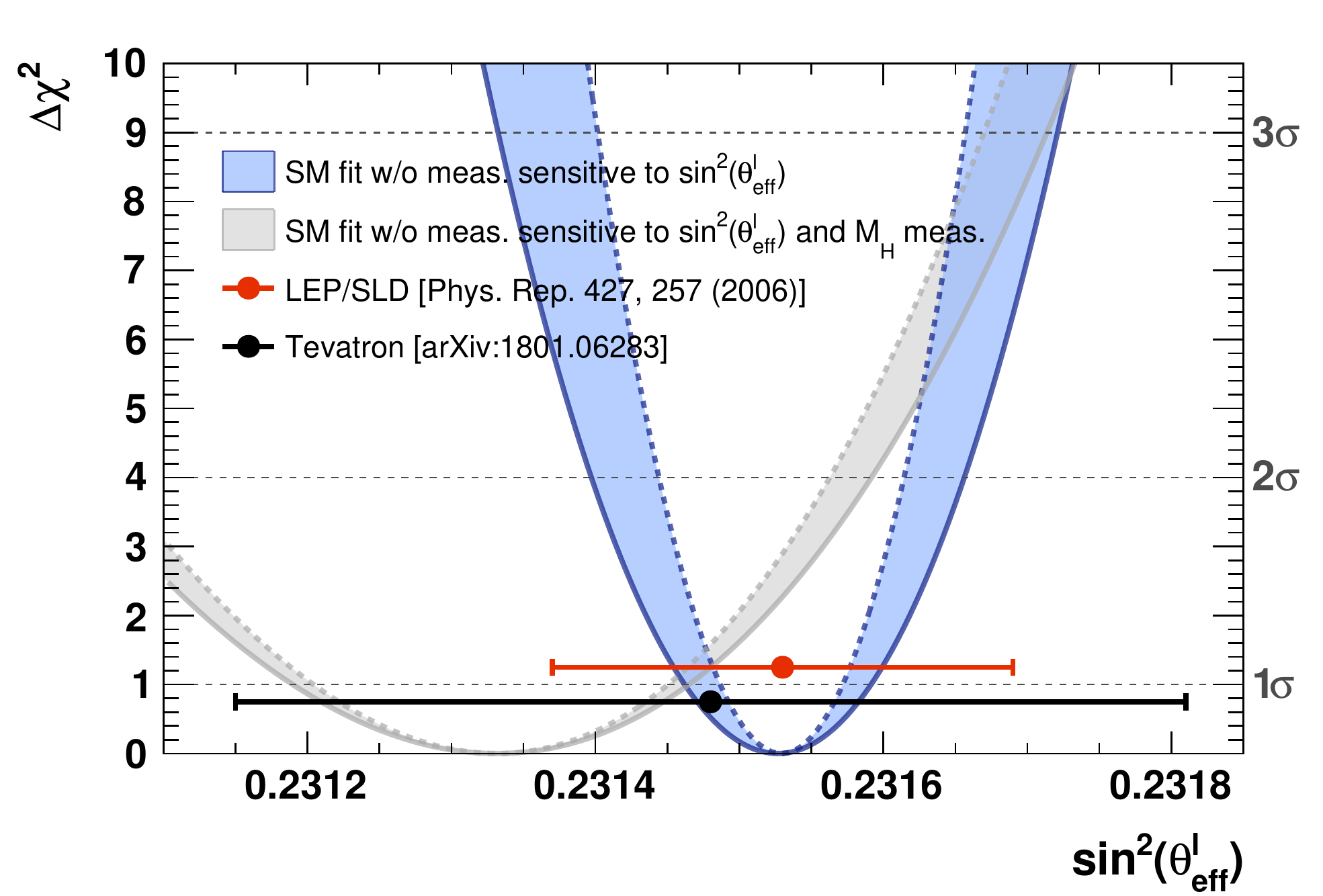}
\includegraphics[width=\defaultDoubleFigureScale\textwidth]{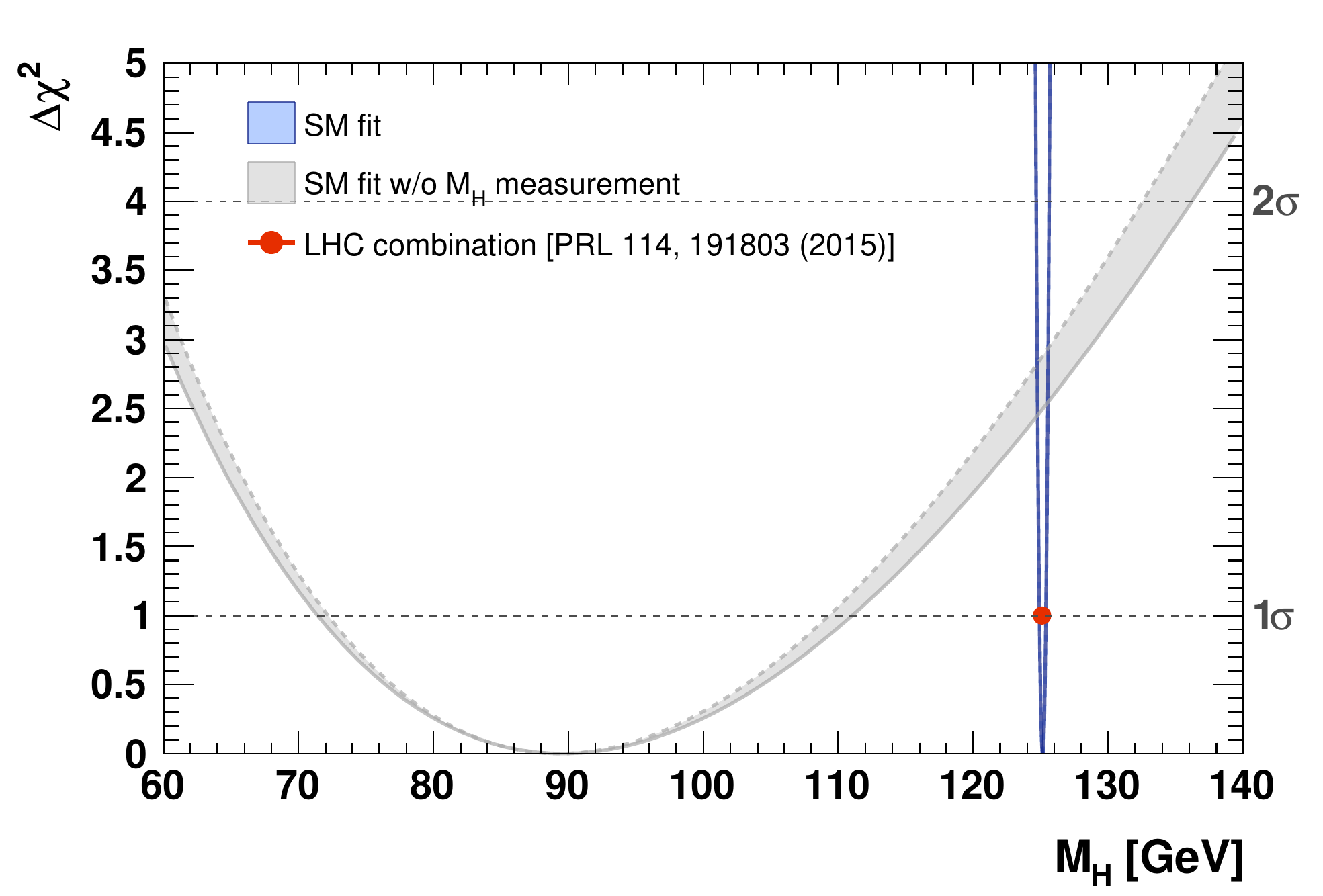}
\includegraphics[width=\defaultDoubleFigureScale\textwidth]{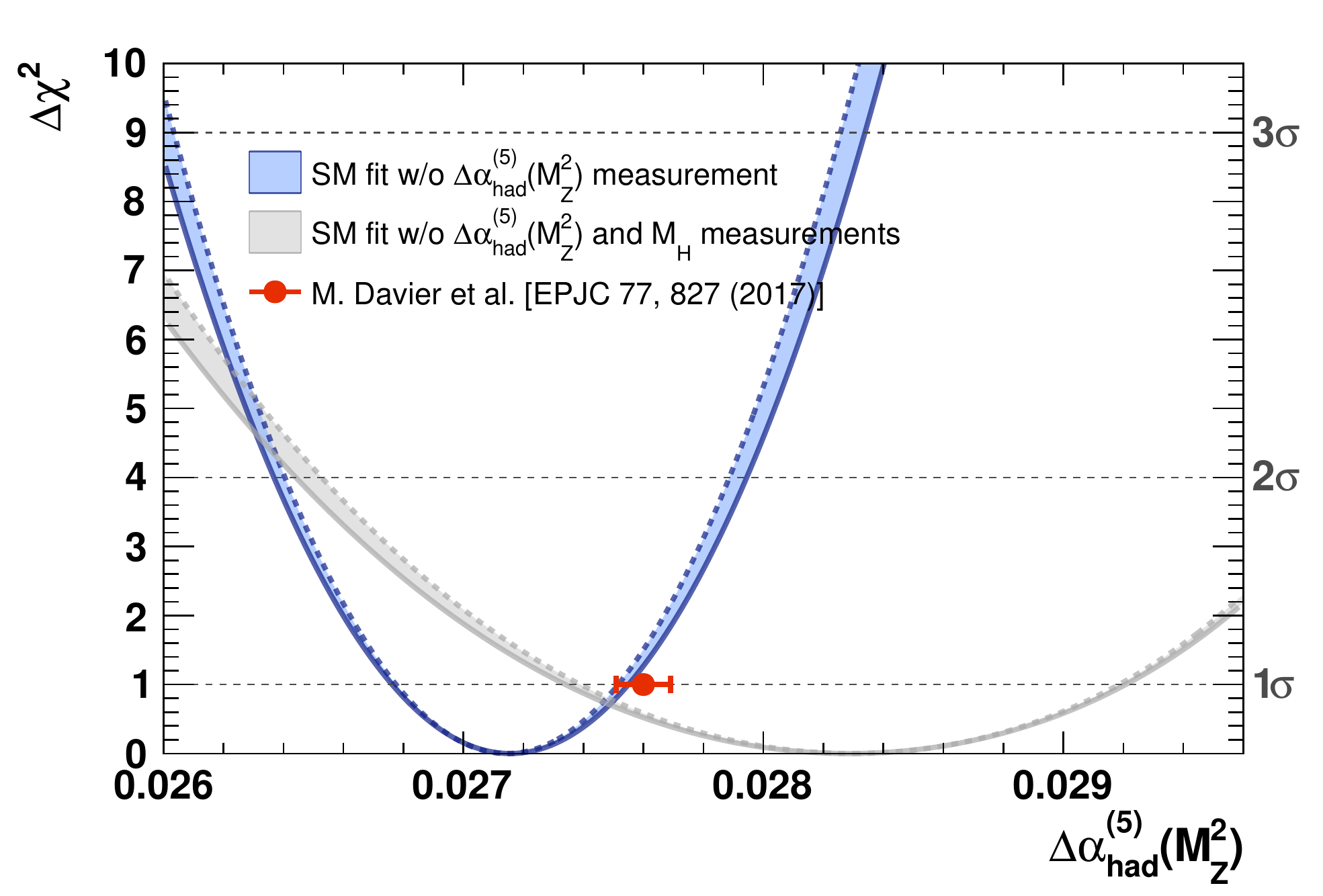}
\includegraphics[width=\defaultDoubleFigureScale\textwidth]{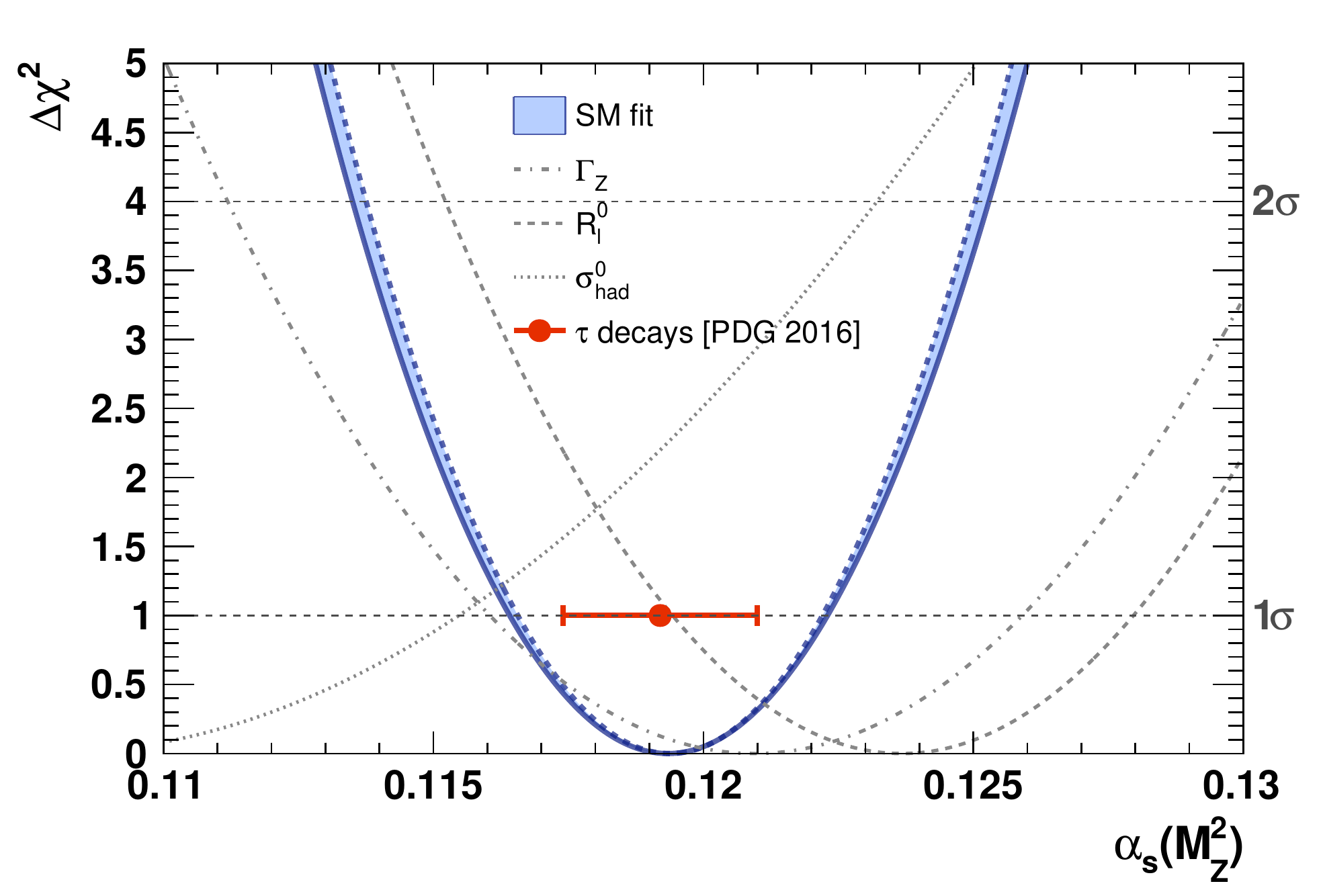}
\end{center}
\vspace{-0.1cm}
\caption[]{Scans of \DeltaChi as a function of $M_W$ (top left),
  $m_t$ (top right), \sinleff (middle left), $M_H$ (middle right),
  $\dalphaHadMZ$ (bottom left)  and $\as(M_{Z}^{2})$ (bottom right), 
  under varying conditions. The results of the  fits without and with the 
  measurement of \MH as input are  shown in grey and blue colours, 
  respectively. The solid and dotted lines represent the results when
  including or excluding the theoretical uncertainties. 
  The data points with uncertainty bars indicate the 
  direct measurements of a given observable. }
\label{fig:scans}
\end{figure}

Figure~\ref{fig:scans} displays \DeltaChi fit profiles for the indirect 
determination of some of the electroweak observables.\footnote{The indirect determination 
profiles are obtained by excluding the input measurement of the respective observable 
from the fit (see figure legends).}
The results are shown for fits
including (blue) and excluding (grey) the direct $M_H$ measurement
highlighting the strong impact of the $M_H$ measurement on the fit
constraints. The direct measurement of each observable with its $1\sigma$ uncertainty 
are indicated by the data points at $\Delta\chi^2=1$. The detailed predictions of the 
fit are given in Table~\ref{tab:results}. 

The fit indirectly determines the $W$ mass to be
\beqn
  M_W &=& 80.3535 
          \pm 0.0027_{m_t} \pm 0.0030_{\delta_{\rm theo} m_t} \pm 0.0026_{M_Z} \pm 0.0026_{\as} \nonumber \\ 
      & & \phantom{80.3535} 
           \pm 0.0024_{\Delta\alpha_{\rm had}} \pm 0.0001_{M_H} \pm 0.0040_{\delta_{\rm theo} M_W}\:{\rm GeV}\,, \nonumber \\[0.2cm]
      &=& 80.354 \pm 0.007_{\rm tot} \:{\rm GeV} \,,
\label{eq:mw}
\eeqn
and the effective leptonic weak mixing angle as
\beqn
  \sinleff &=& 0.231532
                 \pm 0.000011_{m_t} \pm 0.000016_{\delta_{\rm theo} m_t} \pm 0.000012_{M_Z} \pm 0.000021_{\as} \nonumber \\
           & & \phantom{0.231508}
                  \pm 0.000035_{\Delta\alpha_{\rm had}} \pm 0.000001_{M_H} \pm 0.000040_{\delta_{\rm theo} \sinleff} \,, \nonumber \\[0.2cm]
      &=& 0.23153 \pm 0.00006_{\rm tot} \:.
\label{eq:sin2t}
\eeqn
When evaluating $\sinleff$ through the parametric formula from Ref.~\cite{Awramik:2006uz}, an 
upward shift of $2\cdot 10^{-5}$ with respect to the fit result is observed, mostly due to the 
inclusion of $M_W$ in the fit. Using the parametric formula the total  
uncertainty is larger by $0.6\cdot 10^{-5}$, as the global fit exploits 
the additional constraint from $M_W$. The fit also constrains 
the nuisance parameter associated with the theoretical uncertainty in the calculation of 
$\sinleff$, resulting in a reduced  theoretical uncertainty 
of $4.0\cdot 10^{-5}$ compared to the $4.7\cdot 10^{-5}$ input uncertainty.

The mass of the top quark is indirectly determined to be 
\beqn
m_t &=& 176.4\pm 2.1\:\gev\,,
\eeqn
with a theoretical uncertainty of 0.6\:\gev induced by the theoretical uncertainty on 
the prediction of $M_W$. The largest potential to improve the precision of the 
indirect determination of $m_t$ is through a more precise measurement of $M_W$. 
Perfect knowledge of $M_W$ would result in an uncertainty on $m_t$ of 0.9\:\gev.

The strong coupling strength at the $Z$-boson mass scale is determined to be
\beqn
\as(M_{Z}^{2}) &=& 0.1194\pm 0.0029\,,
\eeqn
which corresponds to a determination at full next-to-next-to leading order (NNLO) for 
electroweak and strong contributions, and partial strong next-to-NNLO (NNNLO) corrections. 
The theory uncertainty of this result is $0.0009$, which is 
shared in equal parts between missing higher orders in the calculations of the radiator 
functions and the partial widths of the $Z$ boson. 
The most important constraints on $\as(M_{Z}^{2})$ come from the measurements of 
$R^{0}_{\l}$, $\Gamma_{Z}$ and $\sigma_{\rm had}^{0}$, also shown in Fig.~\ref{fig:scans}. 
The values of $\as(M_{Z}^{2})$ obtained 
from the individual measurements are $0.1237\pm0.0043$ ($R^{0}_{\l}$), 
$0.1209\pm0.0049$ ($\Gamma_{Z}$) and $0.1078\pm0.0076$ ($\sigma_{\rm had}^{0}$). 
A fit to all three measurements results in a value of $0.1203 \pm 0.0030$, which is 
only slightly less precise than the result of the full fit. 
The results obtained for $\as(M_{Z}^{2})$ are stable with respect to additional 
invisible beyond-the-standard-model contributions to $\Gamma_{Z}$.

No significant deviation from the direct measurements is observed in 
any of these predictions.
The indirect determinations of $M_W$ and $\sinleff$ outperform the
direct measurements in precision while the indirect determinations of $m_t$ and
$\as(M_{Z}^{2})$ are competitive to other experimental results. 

\subsubsection*{Oblique parameters}

\begin{figure}
\begin{center}
\includegraphics[width=\defaultSingleFigureScale\textwidth]{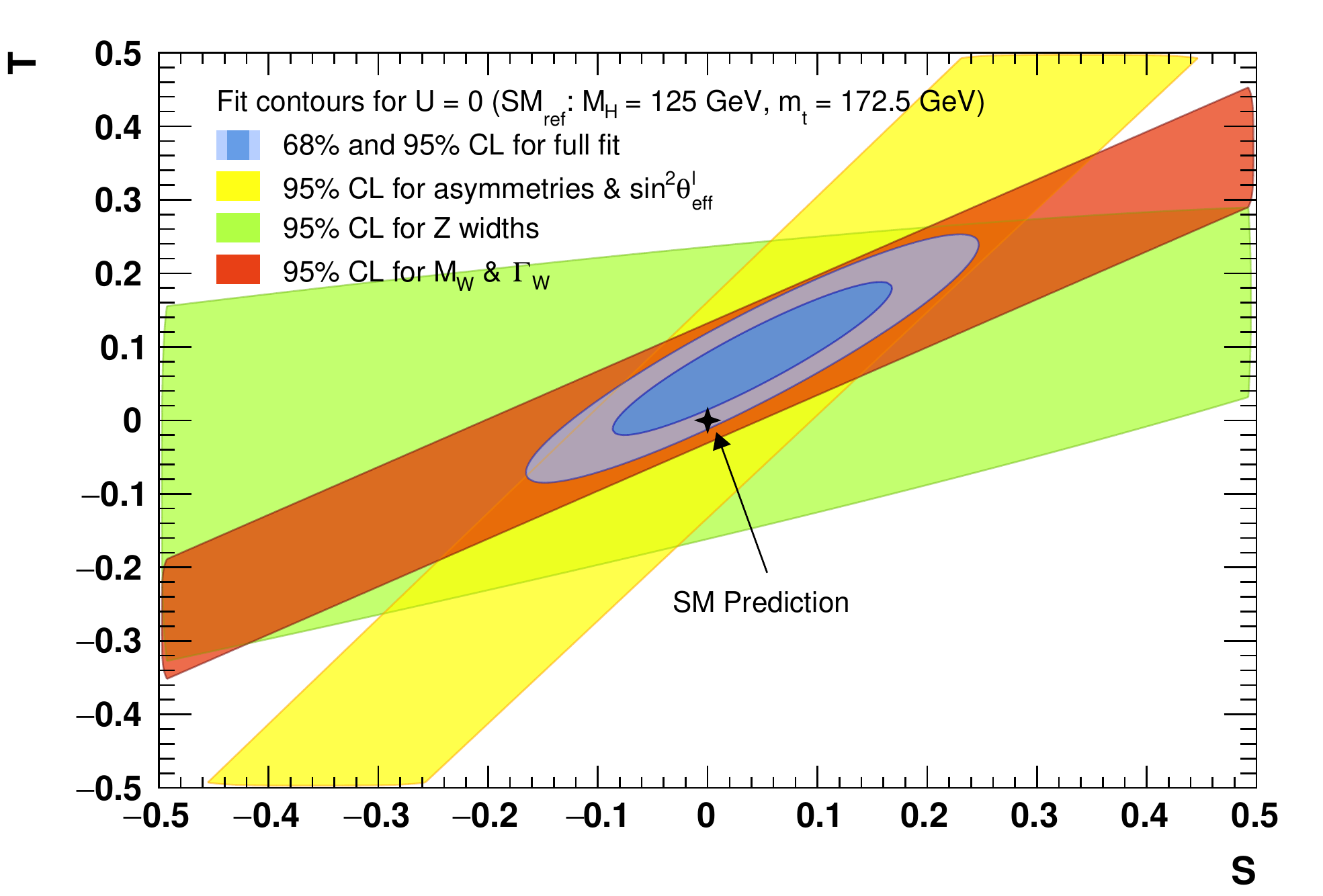}
\end{center}
\vspace{-0.3cm}
\caption[]{Constraints in the oblique parameters $S$ and $T$, with the $U$ parameter fixed to zero, 
using all observables (blue). Individual constraints are shown from the asymmetry and 
direct \sinleff measurements (yellow), the $Z$ partial and total widths (green) 
and $W$ mass and width (red), with confidence levels drawn for one degree of freedom. 
\label{fig:STU}}
\end{figure}

Using the updated SM reference values $M_{H,{\rm ref}}=125$\:GeV and $m_{t,{\rm
    ref}}=172.5$\:GeV we obtain for the {\it oblique parameters}
denoted $S$, $T$, $U$ ~\cite{Peskin:1990zt,Peskin:1991sw} the following values:
\beq
  S= \SParam\,, \hspace{0.5cm}
 T= \TParam\,, \hspace{0.5cm}
  U=\UParam\,,
\label{eq:stu}
\eeq 
with correlation coefficients of $\STParamCor$ between $S$ and $T$, $\SUParamCor$ ($\TUParamCor $) between $S$
and $U$ ($T$ and $U$).
Fixing $U=0$ one obtains $S|_{U=0}= \SParamNU$ and $T|_{U=0}= \TParamNU$, with a correlation coefficient of
$\STParamCorNo$.
The constraints on $S$ and $T$ for a fixed
value of $U=0$ are shown in Fig.~\ref{fig:STU}.

\pagebreak

\section{Global fits in the two-Higgs-doublet model}
\label{sec:2hdm}

Combining information from the electroweak precision data, Higgs boson coupling
measurements, flavour observables and the anomalous magnetic moment of the muon
we derive in this section constraints  on parameters of various 2HDM scenarios.

Besides the four mass parameters for the scalars, $M_{h}$,
$M_{H}$, $M_{A}$, and $M_{H^{\pm}}$, the 2HDM introduces the angle $\alpha$, 
which describes the mixing of the two neutral Higgs fields $h$ and $H$, 
and the angle $\beta$ that fixes the ratio of the vacuum expectation values 
of the two Higgs doublets, $\tanb=v_2/v_1$. We only
consider 2HDM scenarios with a $\mathbb{Z}_2$
symmetric potential with a dimension-two softly broken term
proportional to the scale parameter $M_{12}^2$.

Depending on the Yukawa couplings of the two Higgs doublets, the 2HDM may 
introduce dangerous flavour-changing neutral currents (FCNCs) and CP 
violating interactions. CP conservation can be maintained by fixing the 
Higgs boson couplings for up-type quarks, down-type quarks, and leptons to
specific values~\cite{Haber:1978jt,Branco:2011iw}.  
In this work, four CP conserving 2HDM scenarios are studied. In the {\em Type-I} scenario, 
only one of the two Higgs doublets is allowed to couple to fermions, while the 
other couples to the gauge bosons. The {\em Type-II} scenario is defined by a 
separation of the Yukawa interactions: one Higgs doublet couples only to up-type 
quarks and the other only to down-type quarks and charged leptons. The Type-II 
2HDM resembles the Higgs sector in the Minimal Supersymmetric Standard Model. 
The third, {\em lepton specific} scenario is similar to the Type-I model with the 
difference that leptons only couple to the other Higgs doublet that does not 
interact with the quarks. Finally, the fourth, {\em flipped} scenario is the same 
as the Type-II model with swapped lepton couplings to the Higgs doublets. 

Throughout this section the lightest scalar Higgs boson, $M_{h}$, is identified 
with the observed Higgs boson with mass fixed to $125.09\pm0.24$~GeV~\cite{Aad:2015zhl}. 
If not stated otherwise, all other 2HDM model parameters are allowed to vary 
within the intervals: $130<M_{H},M_{A}<1000\:\gev$, $100<M_{H^{\pm}}<1000\:\gev$,
$0 \le \beta-\alpha \le \pi$, $0.001 < \tanb < 50$,
and $-8\cdot10^5< M_{12}^2<8\cdot10^5\:\gev^2$. No contribution from new physics 
other than the 2HDM is assumed.

Direct searches for additional Higgs bosons in collider experiments can be 
interpreted in the context of the 2HDM (see, for example, Ref.~\cite{Arbey:2017gmh}).
However, due to the large freedom in the choice of the 2HDM parameters, these search results 
provide  only weak absolute exclusion limits on the masses of the scalars.
From searches for a charged Higgs boson by the LEP experiments~\cite{Abbiendi:2013hk}
a lower limit of $M_{H^{\pm}}>72.5\:\gev$ was reported for the Type-I scenario, while
a limit of $M_{H^{\pm}}\gtrsim150\:\gev$ can be derived from searches at the LHC for 
the Type-II scenario~\cite{Arbey:2017gmh}. Stronger mass limits mainly on $M_{H^{\pm}}$ 
can be obtained for specific regions of $\tanb$.

\subsection{Constraints from Higgs boson coupling measurements}

A second Higgs doublet modifies the coupling strengths of the lightest 
neutral Higgs boson $h$ to SM particles compared to those of the SM Higgs boson.
The modifications depend on the 2HDM scenario and parameters in particular 
the angles $\alpha$ and $\beta$. Constraints on $h$ are derived from the joint 
ATLAS and CMS Higgs boson coupling analysis~\cite{Khachatryan:2016vau} 
in which measurements sensitive to 
five Higgs boson production modes (ggF, VBF, $WH$, $ZH$, $t\overline{t}H$) 
and five decay modes ($\gamma\gamma$, $WW$, $ZZ$, $\tau\tau$, $b\overline{b}$) were 
combined. 
We make use of the relative signal strengths $\mu_{ij}$ defined as the ratio
of measured over predicted cross section times branching ratio, 
$\mu_{ij}=(\sigma_i\cdot\BR_j)/(\sigma_i^{\rm SM}\cdot\BR_j^{\rm SM})$.
We include the 20 (out of the 25 possible) $\mu_{ij}$ parameters determined by ATLAS and CMS together with their uncertainties 
and correlations. 
A validation of our results is discussed in the Appendix on 
page~\pageref{appendix}. 

The corresponding SM predictions and uncertainties are taken from 
Ref.~\cite{Heinemeyer:2013tqa}. The signal strength measurements are compared
with the theory predictions for the 2HDM calculated with the program 
\texttt{2HDMC}~\cite{Eriksson:2009ws}.\footnote{\texttt{2HDMC} computes the couplings 
of all five Higgs boson states to SM particles for a given set of parameters in a 
CP conserving 2HDM with general Yukawa structure. From these couplings, production 
and decay rates of the Higgs boson states can be derived. Most decay widths are 
calculated at leading QCD order in \texttt{2HDMC}. Higher order QCD corrections 
are included for couplings to fermion and gluon pairs.}
In the calculation of the $\mu_{ij}$ for the 2HDM also the denominator $\sigma_i^{\rm SM}\cdot\BR_j^{\rm SM}$
is determined using \texttt{2HDMC} for consistency.
Since more precise theory predictions for the SM cross sections and branching ratios exist and are used 
for the normalisation of the results in~\cite{Khachatryan:2016vau},
theory uncertainties in the SM prediction from~\cite{Heinemeyer:2013tqa} are taken into account as additional
scaling (nuisance) parameters of the $\mu_{ij}$.

\begin{figure}[t]
\begin{center}
\includegraphics[width=\defaultDoubleFigureScale\textwidth]{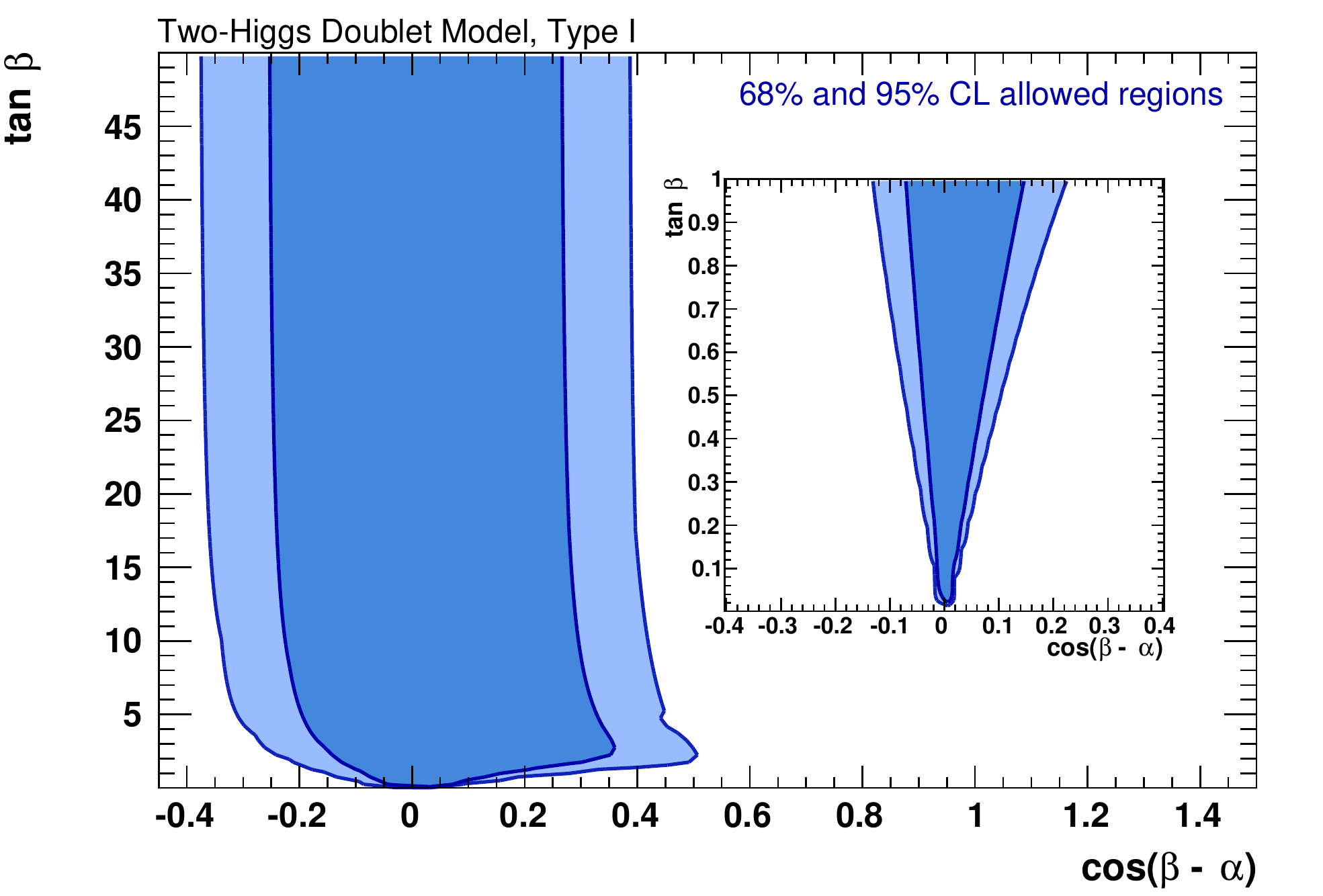}
\includegraphics[width=\defaultDoubleFigureScale\textwidth]{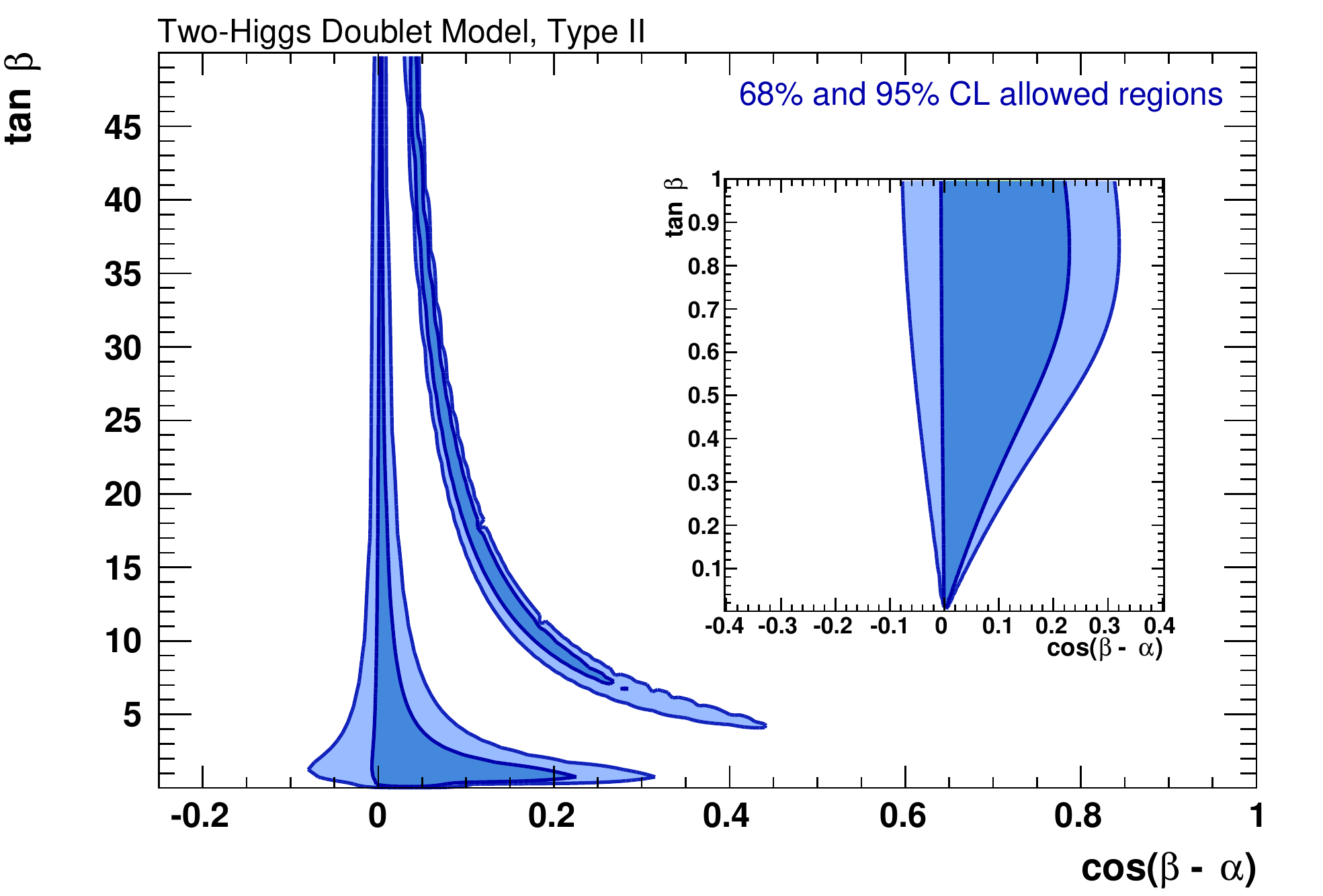}
\includegraphics[width=\defaultDoubleFigureScale\textwidth]{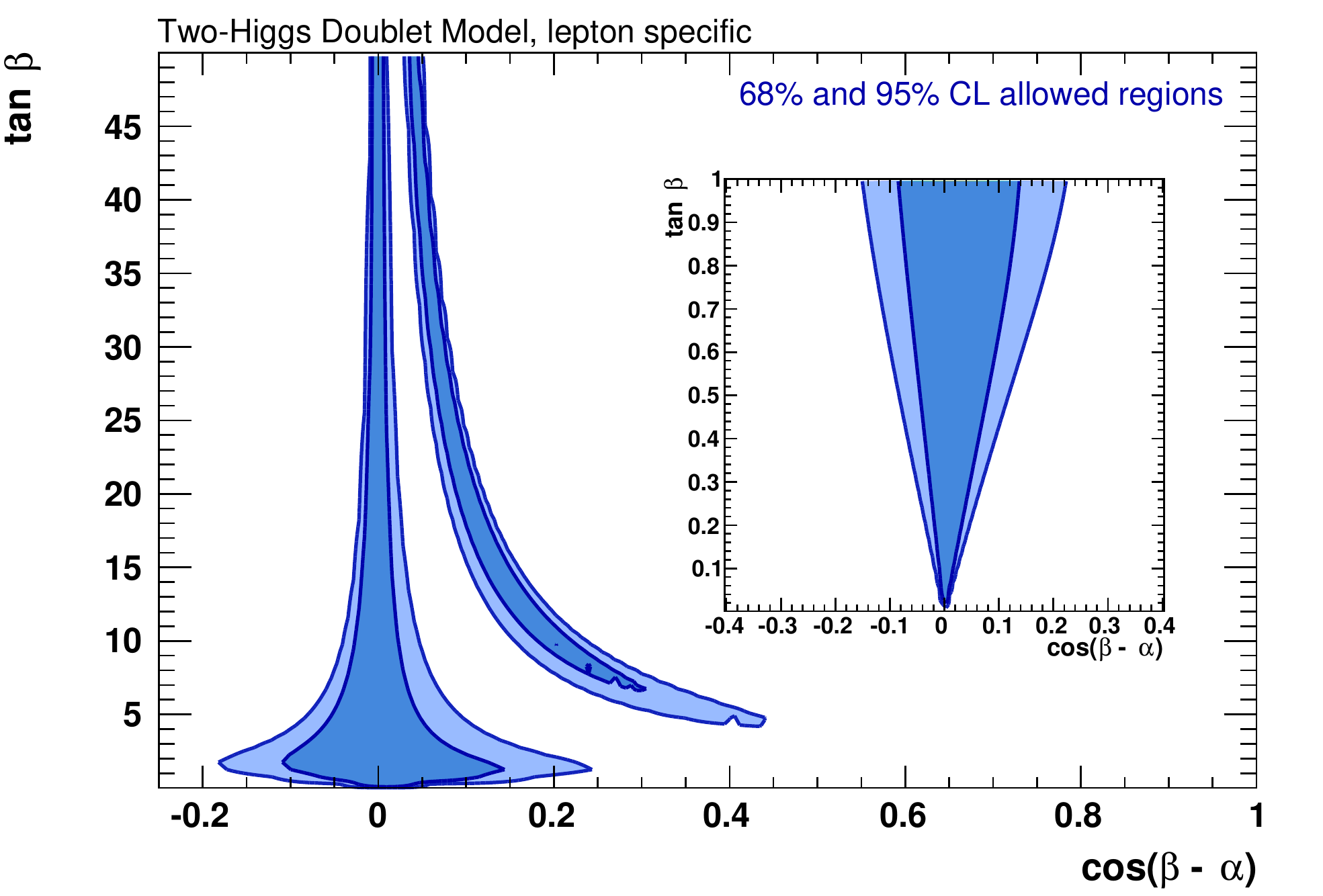}
\includegraphics[width=\defaultDoubleFigureScale\textwidth]{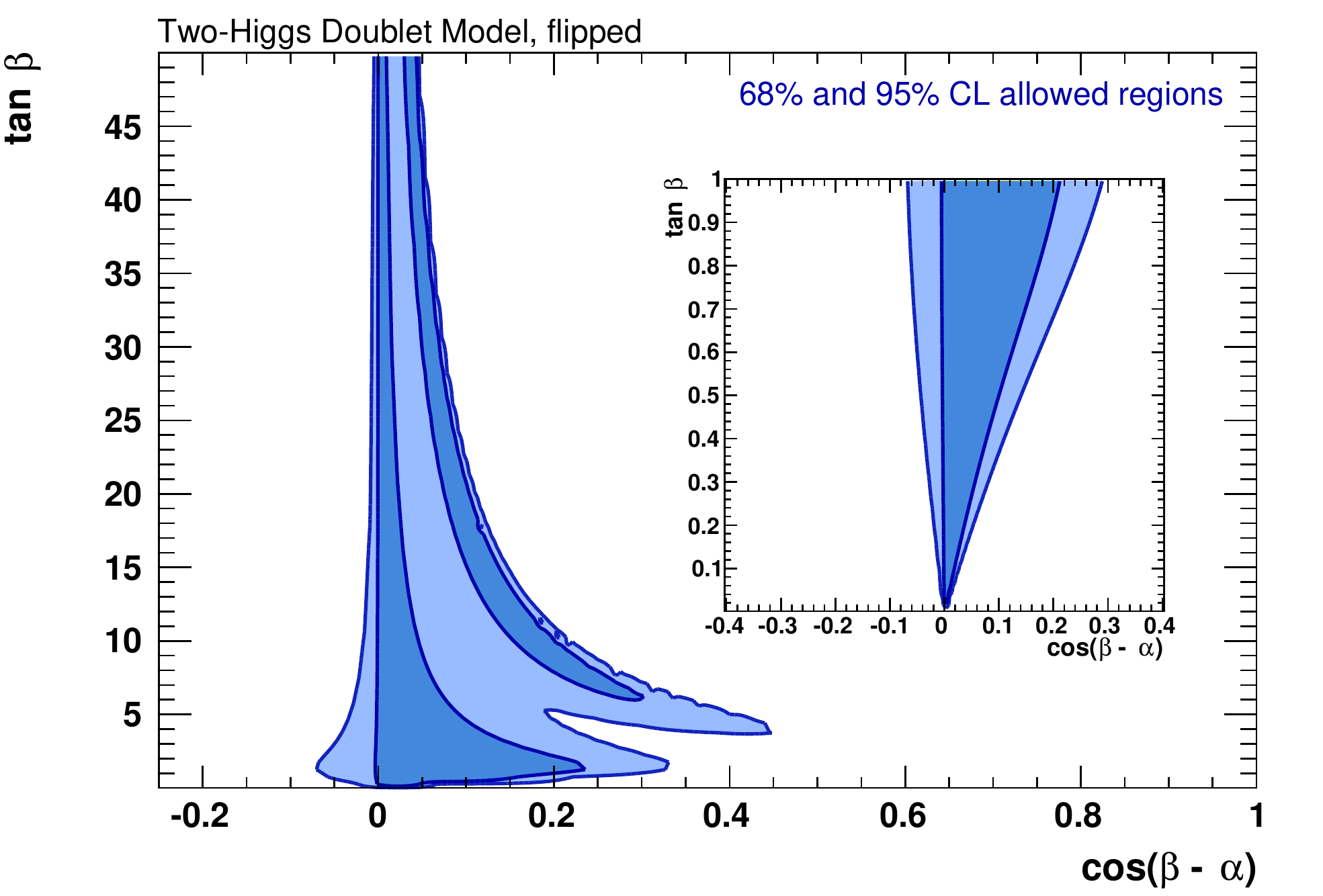}
\vspace{-0.8cm}
\end{center}
\caption[]{Results from 2HDM fits using the ATLAS and CMS combined Higgs coupling 
  strength measurements. Shown are allowed parameter regions (68\% and 95\% CL) for the four 
  2HDM scenarios from scans of $\tanb$ versus $\cos(\beta-\alpha)$: 
  Type-I (top left), Type-II (top right), lepton specific (bottom left) and 
  flipped (bottom right) 2HDMs. The figure insets show a zoom of the
  region with $\tanb<1$.
}
\label{fig:bma_tanb}
\end{figure}

The constraints from the Higgs boson signal strength measurements on the 
four 2HDM scenarios are shown as 68\% and 95\% CL allowed regions
in the $\tanb$ versus $\cos(\beta-\alpha)$
plane in Fig.~\ref{fig:bma_tanb}.\footnote{Theoretical 
  bounds from positivity of the Higgs potential, tree-level unitarity, and 
  perturbativity of the quartic Higgs boson couplings as implemented in 
  \texttt{2HDMC} were found to give no additional constraints in these  
  figures.}   

The angles $\alpha$ and $\beta$ are highly constrained in all 2HDM scenarios 
except for Type-I. The allowed parameter regions are concentrated in two bands 
corresponding to solutions with $\beta\pm\alpha=\pi/2$. For $\beta-\alpha=\pi/2$, 
the Yukawa structure of the SM is reproduced ({\it alignment limit}). The case $\beta+\alpha=\pi/2$ 
differs from the SM-like Yukawa couplings by a sign flip that is still allowed by the 
combined coupling strengths measurements. These constraints are differently 
pronounced in the four 2HDM scenarios as they depend on the Yukawa coupling strengths. 
In the Type-I scenario (top left panel in Fig.~\ref{fig:bma_tanb}) the Yukawa couplings 
of $h$ to all fermions are proportional to $\cos\alpha/\sin\beta$. The constraints 
are stronger in the other three scenarios as the Yukawa coupling for at least 
one fermion type is proportional to $-\sin\alpha/\cos\beta$. In the flipped scenario 
(bottom right panel) only the Yukawa coupling to down-type quarks is given by 
$-\sin\alpha/\cos\beta$, which is constrained by the measurements of 
$H\to{b\overline{b}}$. Measurements of $H\to\tau^+\tau^-$ give stronger bounds
in the Type-II (top right panel) and lepton specific (bottom left panel) scenarios 
where the Yukawa couplings to leptons is given by $-\sin\alpha/\cos\beta$.
In all scenarios, the measurements of Higgs boson decays to $W$ and $Z$ boson pairs
disfavour large values of $\cos(\beta-\alpha)$.
Similar constraints have  been obtained by the ATLAS collaboration~\cite{Aad:2015pla}.

\subsection{Constraints from flavour observables}

Because tree-level FCNC transitions are forbidden by construction in the four 2HDM 
scenarios considered, flavour violation only arises at loop level by the exchange of a 
charged Higgs boson with observable 
strength depending on the parameters $M_{H^{\pm}}$ and $\tanb$.

\subsubsection*{Experimental input data and theory calculation}

\begin{table}[t]
\begin{tabular*}{\textwidth}{@{\extracolsep{\fill}}lcc} 
\hline\noalign{\smallskip}
Observable & Value & Reference \\ 
\noalign{\smallskip}\hline\noalign{\smallskip}
$\BR(B\rightarrow X_s\gamma)$ for $E_{\gamma}>1.6$~GeV & $(3.32\pm0.15_{\rm stat+syst})\cdot10^{-4}\pm7\%_{\rm theo}$ & \cite{HFLAV,Misiak:2006zs,Misiak:2015xwa}\\ 
\noalign{\smallskip}\hline\noalign{\smallskip}
$R(D)$ & $0.407\pm0.039_{\rm stat}\pm0.024_{\rm syst}\pm0.008_{\rm theo}$ & ~\cite{HFLAV,Aoki:2016frl} \\
\noalign{\smallskip}
$R(D^{*})$ & $0.304\pm0.013_{\rm stat}\pm0.007_{\rm syst}\pm0.003_{\rm theo}$ & \cite{HFLAV,Fajfer:2012vx}\\ 
\noalign{\smallskip}\hline\noalign{\smallskip}
$\BR(B\rightarrow\tau\nu)$ & $(1.06\pm0.19)\cdot10^{-4}$ & \cite{HFLAV} \\ 
\noalign{\smallskip}\hline\noalign{\smallskip}
$\BR(B_s\rightarrow\mu\mu)$ (CMS) & $(2.8^{\,+1.0}_{\,-0.9})\cdot10^{-9}$ & \cite{CMS:2014xfa} \\ 
$\BR(B_s\rightarrow\mu\mu)$ (LHCb)& $(3.0\pm{0.6_{\rm stat}}^{\,+0.3}_{\,-0.2_{\rm syst}})\cdot10^{-9}$ & \cite{Aaij:2017vad} \\ 
\noalign{\smallskip}\hline\noalign{\smallskip}
$\BR(B_d\rightarrow\mu\mu)$ (CMS)& $(4.4^{\,+2.2}_{\,-1.9})\cdot10^{-10}$ & ~\cite{CMS:2014xfa} \\ 
$\BR(B_d\rightarrow\mu\mu)$ (LHCb) & $(1.5^{\,+1.2}_{\,-1.0_{\rm stat}}{}^{\,+0.2}_{\,-0.1_{\rm syst}})\cdot10^{-10}$ & ~\cite{Aaij:2017vad} \\ 
\noalign{\smallskip}\hline\noalign{\smallskip}
$\BR(D_s\rightarrow\mu\nu)$ & $(5.54\pm0.20_{\rm stat}\pm0.13_{\rm syst})\cdot10^{-3}$ & \cite{HFLAV}\\
$\BR(D_s\rightarrow\tau\nu)$ & $(5.51\pm0.18_{\rm stat}\pm0.16_{\rm syst})\cdot10^{-2}$ & \cite{HFLAV}\\
\noalign{\smallskip}\hline\noalign{\smallskip}
$\Delta m_{d}$  & $(0.5065\pm0.0019)$~ps$^{-1}$ & \cite{HFLAV}\\
$\Delta m_{s}$  & $(17.757\pm0.021)$~ps$^{-1}$ & \cite{HFLAV}\\ 
\noalign{\smallskip}\hline\noalign{\smallskip}
$\BR(K\rightarrow \mu\nu)/\BR(\pi\rightarrow \mu\nu)$ & $0.6357\pm0.0011$ & \cite{Olive:2016xmw}\\
\noalign{\smallskip}\hline
\end{tabular*}
\caption{Flavour physics observables and values used in the 2HDM fit.}
\label{tab:flav_obs}
\end{table}

The flavour physics observables taken into account in our analysis are listed in 
Table~\ref{tab:flav_obs} and briefly described below. 

For the branching fraction of the radiative decay 
$\BR(B\rightarrow X_s\gamma)$ with $E_{\gamma}>1.6$~GeV we use the value of
the Heavy Flavour Averaging Group (HFLAV)~\cite{HFLAV} which combines 
measurements from the BABAR~\cite{Aubert:2007my,Lees:2012ym,Lees:2012wg}, 
Belle~\cite{Limosani:2009qg,Saito:2014das,Belle:2016ufb}, and CLEO~\cite{Chen:2001fja} 
experiments. The prediction for $\BR(B\rightarrow X_s\gamma)$ has been 
adopted from Ref.~\cite{Misiak:2015xwa} and includes QCD corrections up 
to NNLO~\cite{Hermann:2012fc}. We make use of the code implementation kindly provided by M.~Misiak.

HFLAV also combined measurements of the semileptonic decay ratios of neutral $B$ mesons 
$R(D^{(*)})=\BR(\Bzb\to D^{(*)+}\tau^-\nub)/\BR(\Bzb\to D^{(*)+}\ell^-\nub)$ 
by BABAR~\cite{Lees:2012xj,Lees:2013uzd}, 
Belle~\cite{Huschle:2015rga,Sato:2016svk,Hirose:2016wfn}, and 
LHCb~\cite{Aaij:2015yra} with a correlation of $-0.23$ between the two 
observables that is taken into account in the fit.
The prediction of $R(D^{(*)})$~\cite{Aoki:2016frl,Fajfer:2012vx,Enomoto:2015wbn} includes 
tree-level contributions of a charged Higgs boson and is based on form 
factors evaluated in Heavy-Quark Effective Theory.
Variations of the parameters $\rho^2_{R(D)}$, $\rho^2_{R(D^*)}$, 
$R_1(1)$, and $R_2(1)$ are included in the fit with
values and correlations taken from Ref.~\cite{HFLAV}.

For the branching ratio $\BR(B\rightarrow\tau\nu)$ we  use  the HFLAV 
average~\cite{HFLAV} of   measurements from BABAR~\cite{Lees:2012ju} 
and Belle~\cite{Kronenbitter:2015kls}.
For the prediction of $\BR(B\rightarrow\tau\nu)$ in the 2HDM we use the calculation 
from Ref.~\cite{Isidori:2006pk}, which contains tree-level contributions of a charged Higgs boson where the 
leading $\tanb$ corrections are resummed to all orders~\cite{Isidori:2006pk}.
The theoretical uncertainties in $|V_{ub}|$ and $f_{B_d}$ (see below) are included. 

The latest measurements of $\BR(B_s\rightarrow\mu\mu)$ and $\BR(B_d\rightarrow\mu\mu)$ 
from  LHCb ~\cite{Aaij:2017vad} are combined in our fits with the 
CMS result~\cite{CMS:2014xfa}, 
assuming them uncorrelated. Their theoretical predictions in the 2HDM include NLO corrections 
given in Refs.~\cite{Li:2014fea,Cheng:2015yfu}. The SM contribution to these observables are known up 
to three-loop level in QCD and include NLO electroweak 
corrections~\cite{Hermann:2013kca,Bobeth:2013tba,Bobeth:2013uxa}.
The  predictions depend on the CKM matrix elements $|V_{tb}|$ and $|V_{ts}|$ or $|V_{td}|$, respectively,
and on the respective hadronic parameters $f_{B_s}$ and $f_{B_d}$. 
Uncertainties in these parameters are taken into account in the fit.

The charged Higgs boson of the 2HDM  contributes to the leptonic decays 
of $D_s$ mesons. For the observables $\BR(D_s\rightarrow\mu\nu)$ and 
$\BR(D_s\rightarrow\tau\nu)$ we use the HFLAV averages~\cite{HFLAV} of 
measurements from BABAR~\cite{delAmoSanchez:2010jg}, Belle~\cite{Zupanc:2013byn}, 
and CLEO~\cite{Onyisi:2009th,Alexander:2009ux,Naik:2009tk}.
For the 2HDM predictions we use the analytic expression for the 2HDM tree-level contribution 
to $\BR(D_s\rightarrow\ell\nu)$ from Ref.~\cite{Akeroyd:2009tn} that allows
us to vary the dependencies on $|V_{cs}|$ and  $f_{D_s}$ in the fit.


The charged Higgs boson also contributes via box diagrams to the mixing of the neutral 
$B_d$ and $B_s$ mesons  altering the mixing frequencies 
$\Delta m_{d}$ and/or $\Delta m_{s}$. 
We use again the HFLAV~\cite{HFLAV} experimental averages for these quantities. 
Their predictions in the 2HDM are obtained from analytic expressions of the full one-loop calculation of 
Refs.~\cite{Chang:2015rva,Enomoto:2015wbn} neglecting small terms proportional 
to $m_{b}^2/M_{W}^2$. The predictions depend on the CKM matrix elements 
$|V_{td}|$ and $|V_{ts}|$, the bag parameters $\hat{B}_{d}$ and $\hat{B}_{s}$, 
and the decay constants  $f_{B_d}$ and $f_{B_s}$, respectively, and the  correction 
factor $\eta_B$.

Finally, the 2HDM contributes  at leading order to the ratio 
$\BR(K\rightarrow \mu\nu)/\BR(\pi\rightarrow \mu\nu)$ for which we use 
a value adopted from Ref.~\cite{Olive:2016xmw}, based on the measurement of 
the kaon decay rates~\cite{Ford:1967zz}, and the 2HDM prediction from Ref.~\cite{Enomoto:2015wbn}.
The ratio involves  the CKM matrix elements $|V_{us}|$ and $|V_{ud}|$, 
 the decay constants $f_K$ and $f_\pi$, and an electromagnetic correction 
$\delta_{\rm EM}^{K/\pi}$.

As input values for the unitarity CKM matrix we use the latest available 
results for the all-orders Wolfenstein parameters $A$, $\lambda$, $\rhobar$,  $\etabar$
from Ref.~\cite{Charles:2004jd,Charles:2015gya,CKMfit}, taking them uncorrelated. 
A fully consistent analysis would 
require a combined fit of the Wolfenstein and 2HDM parameters within the 
2HDM~\cite{Deschamps:2009rh}, which is however beyond the scope of this paper. 
Studies in Ref.~\cite{Enomoto:2015wbn} and by ourselves have shown that the 
numerical impact of the 2HDM on the CKM parameters is modest. 
For the CKM element $|V_{ub}|$, occurring mainly in the prediction of the leptonic 
$B^\pm$ branching fraction, we  take the average of inclusive and exclusive 
measurements~\cite{Bevan:2014iga} instead of the CKM fit prediction to
allow for a more conservative uncertainty in view of the  tension between 
the inclusive and exclusive results. 

The input parameters used in the fit are summarised in Table~\ref{tab:flav_input}.

\begin{table}[t]
\setlength{\tabcolsep}{0.0pc}

\begin{tabular*}{\textwidth}{@{\extracolsep{\fill}}lc|lc} 
\hline\noalign{\smallskip}
Parameter & Value & Parameter & Value \\ 
\noalign{\smallskip}\hline\noalign{\smallskip}
$A$ & $0.8250^{\,+0.0071}_{\,-0.0111}$                & $f_{D_s}$ & $(248.2\pm 0.3_{\rm stat} \pm 1.9_{\rm syst})$~MeV\\
$\lambda$ & $0.22509^{\,+0.00029}_{\,-0.00028}$  & $f_{B_s}$ & $(225.6\pm 1.1_{\rm stat}\pm 5.4_{\rm syst})$~MeV\\
$\rhobar$ & $0.1598^{\,+0.0076}_{\,-0.0072}$       & $f_{B_s}/f_{B_d}$ & $1.205 \pm 0.004_{\rm stat} \pm 0.007_{\rm syst}$\\
$\etabar$ & $0.3499^{\,+0.0063}_{\,-0.0061}$       & $\hat{B}_{s}$& $1.320 \pm 0.017_{\rm stat} \pm 0.030_{\rm syst}$  \\
$|V_{ub}|$ & $0.00395\pm0.00038_{\rm exp}\pm0.00039_{\rm theo}$~~~ & $\hat{B}_{s}/\hat{B}_{d}$ & $1.023 \pm 0.013_{\rm stat} \pm 0.014_{\rm syst}$\\
$\rho^2_{R(D)}$ & $1.128\pm0.033$                  & $\eta_B$ & $0.551 \pm 0.0022$ \\
$\rho^2_{R(D^*)}$ & $1.21\pm0.027$                & $f_{K}/f_{\pi}$ & $1.1952 \pm 0.0007_{\rm stat} \pm 0.0029_{\rm syst}$\\
$R_1(1)$ & $1.404\pm0.032$                           & $\delta_{\rm EM}^{K/\pi}$ & $-0.0070 \pm 0.0018$\\
$R_2(1)$ & $0.854\pm0.020$ & & \\
\noalign{\smallskip}\hline
\end{tabular*}
\caption{Parameters used in the fit to the flavour observables. Most values are taken from 
  latest available version of the CKM fit~\cite{CKMfit}. For the CKM matrix element $|V_{ub}|$ 
  we use the average of inclusive and exclusive measurements~\cite{Bevan:2014iga}, while
  all other CKM matrix elements are calculated from the Wolfenstein parameters. 
  The parameters related to the $R(D^{(*)})$ measurements, 
  $\rho^2_{R(D)}$, $\rho^2_{R(D^*)}$, $R_1(1)$,  $R_2(1)$ are taken from Ref.~\cite{HFLAV}.
  Value and uncertainty for $\delta_{\rm EM}^{K/\pi}$ are taken from Ref.~\cite{Antonelli:2010yf}.}
\label{tab:flav_input}
\end{table}

\subsubsection*{Results}

Since most flavour observables are only sensitive to $M_{H^\pm}$ and $\tanb$, separate scans of 
these parameters are performed for each observable. The other 2HDM parameters are ignored
in these scans, with the exception of $\BR(B_{s/d}\rightarrow\mu\mu)$, 
where in addition $M_{H}$, $M_{A}$, and $M_{12}^2$ are allowed to float freely within the bounds 
defined in the introduction of Section~\ref{sec:2hdm} as these two observables 
depend at NLO level on these parameters. In all fits the CKM matrix elements and the 
other parameters given in Table~\ref{tab:flav_input} are allowed to vary within 
their uncertainties.

\begin{figure}[t]
\begin{center}
\includegraphics[width=\defaultDoubleFigureScale\textwidth]{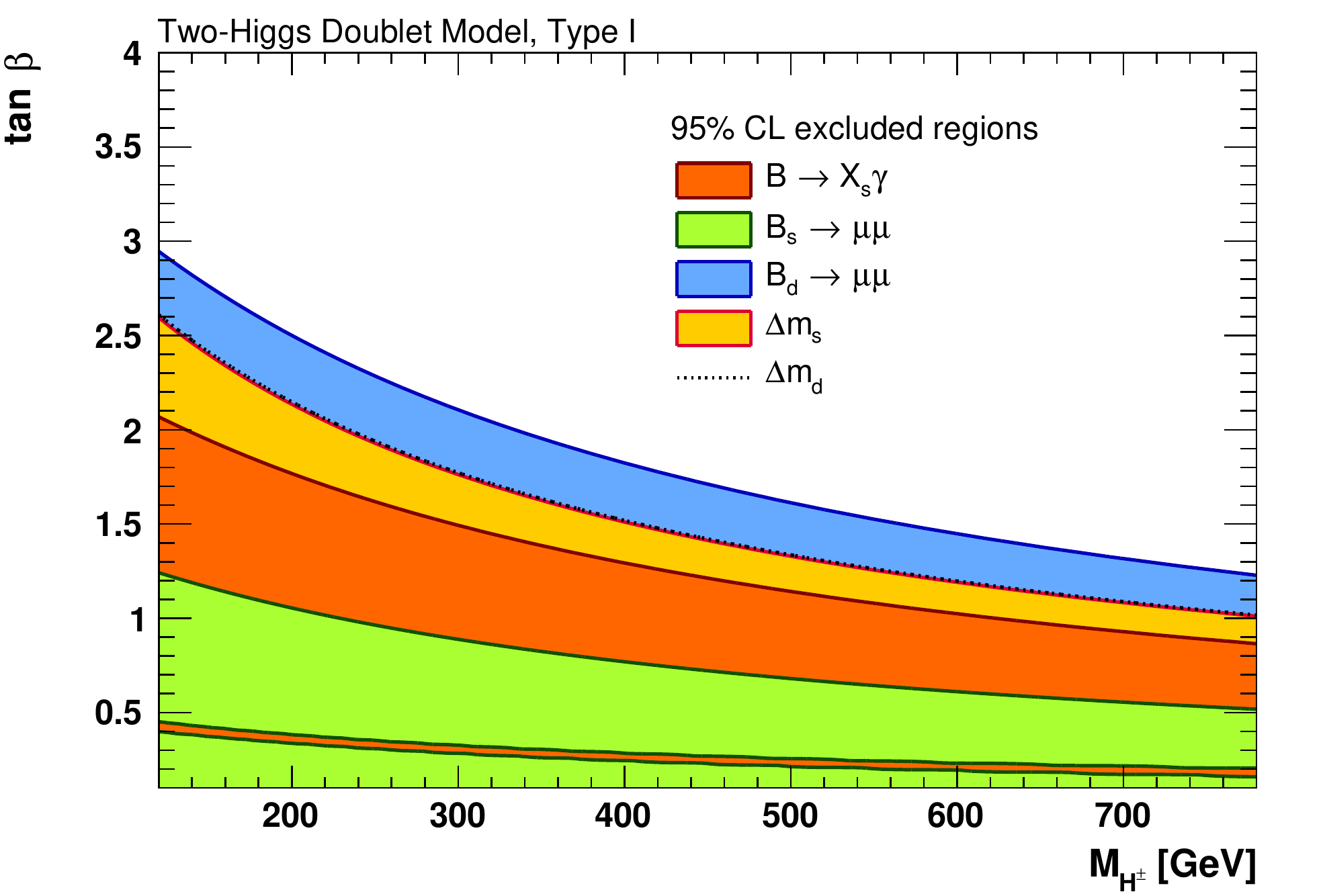}
\includegraphics[width=\defaultDoubleFigureScale\textwidth]{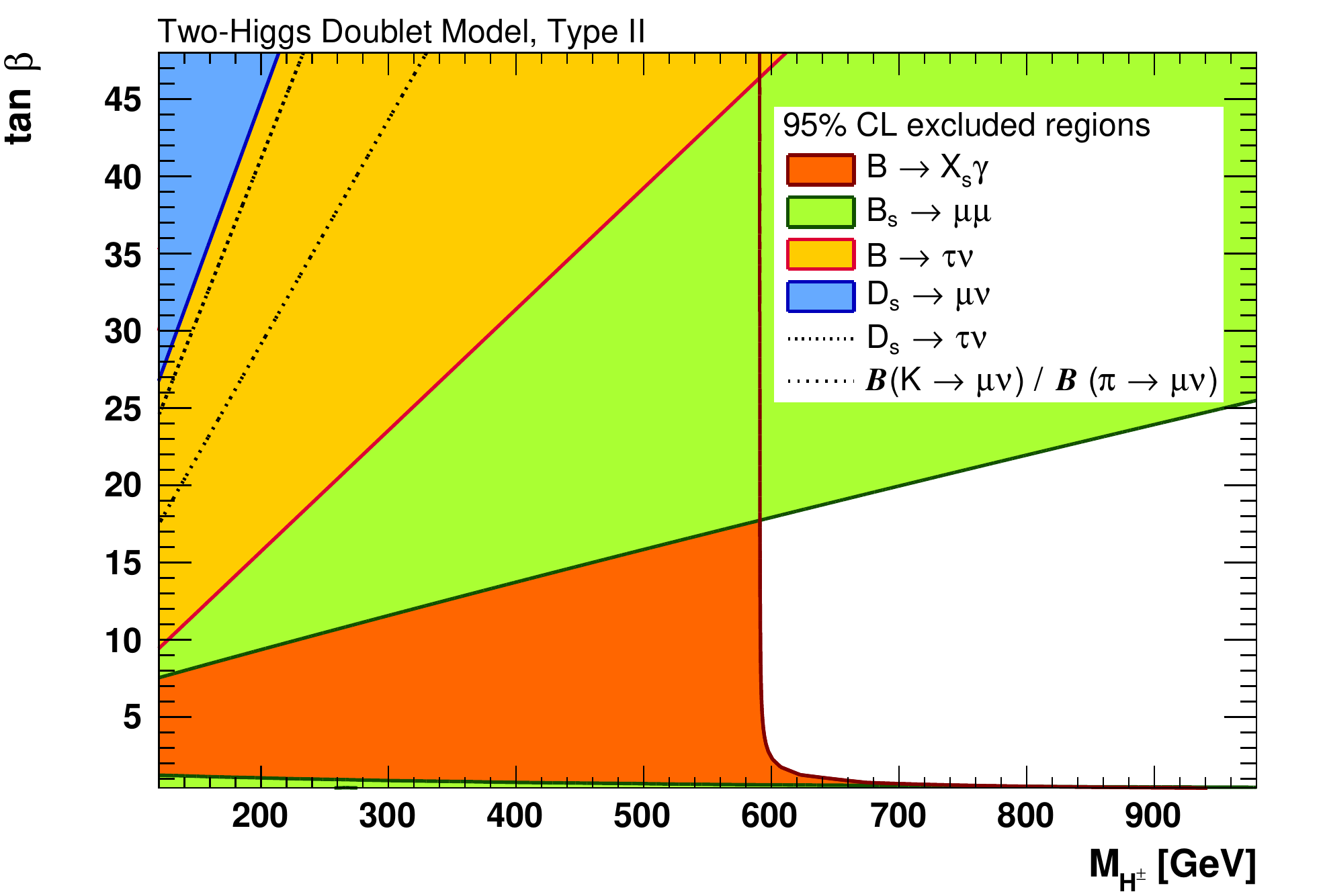}
\includegraphics[width=\defaultDoubleFigureScale\textwidth]{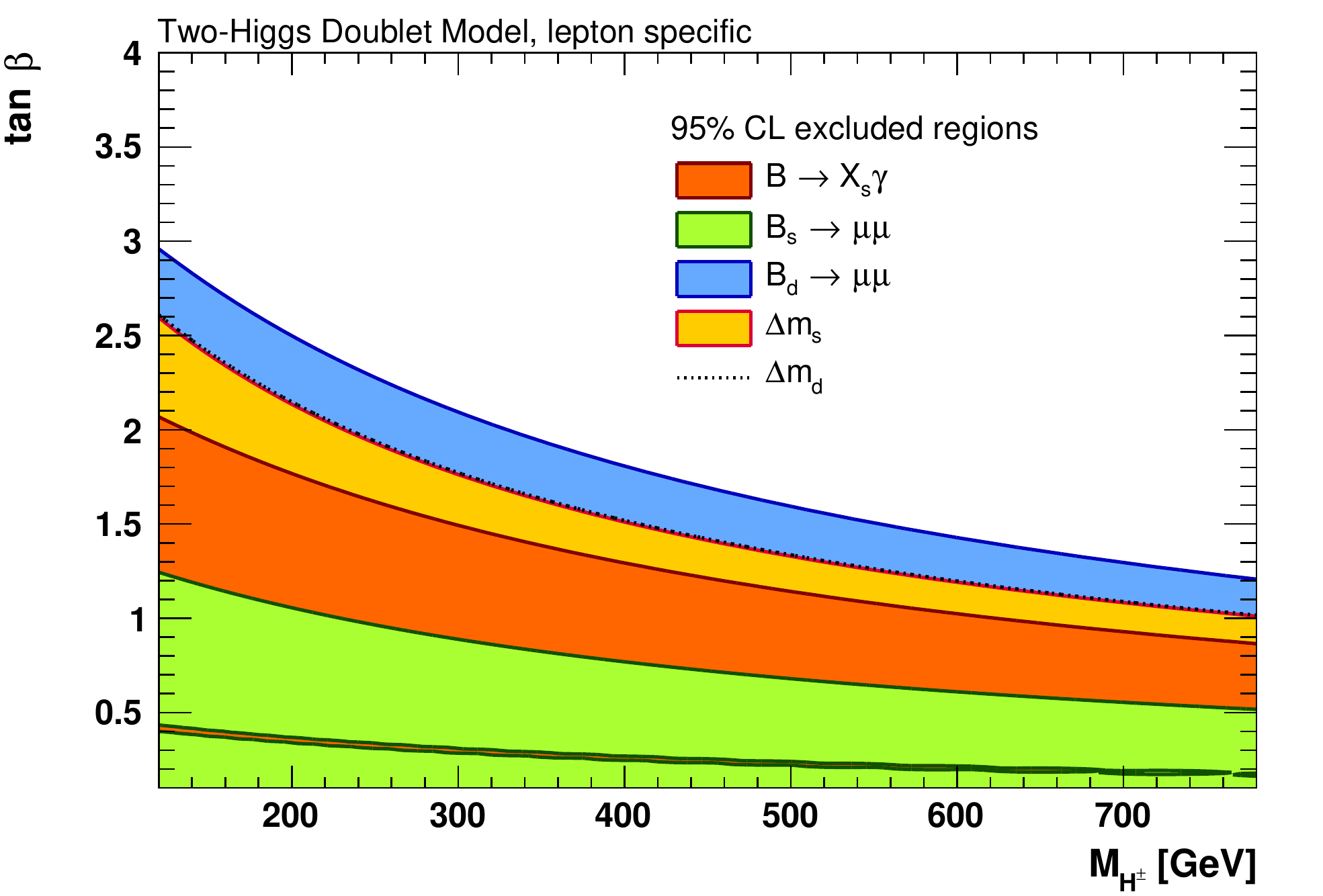}
\includegraphics[width=\defaultDoubleFigureScale\textwidth]{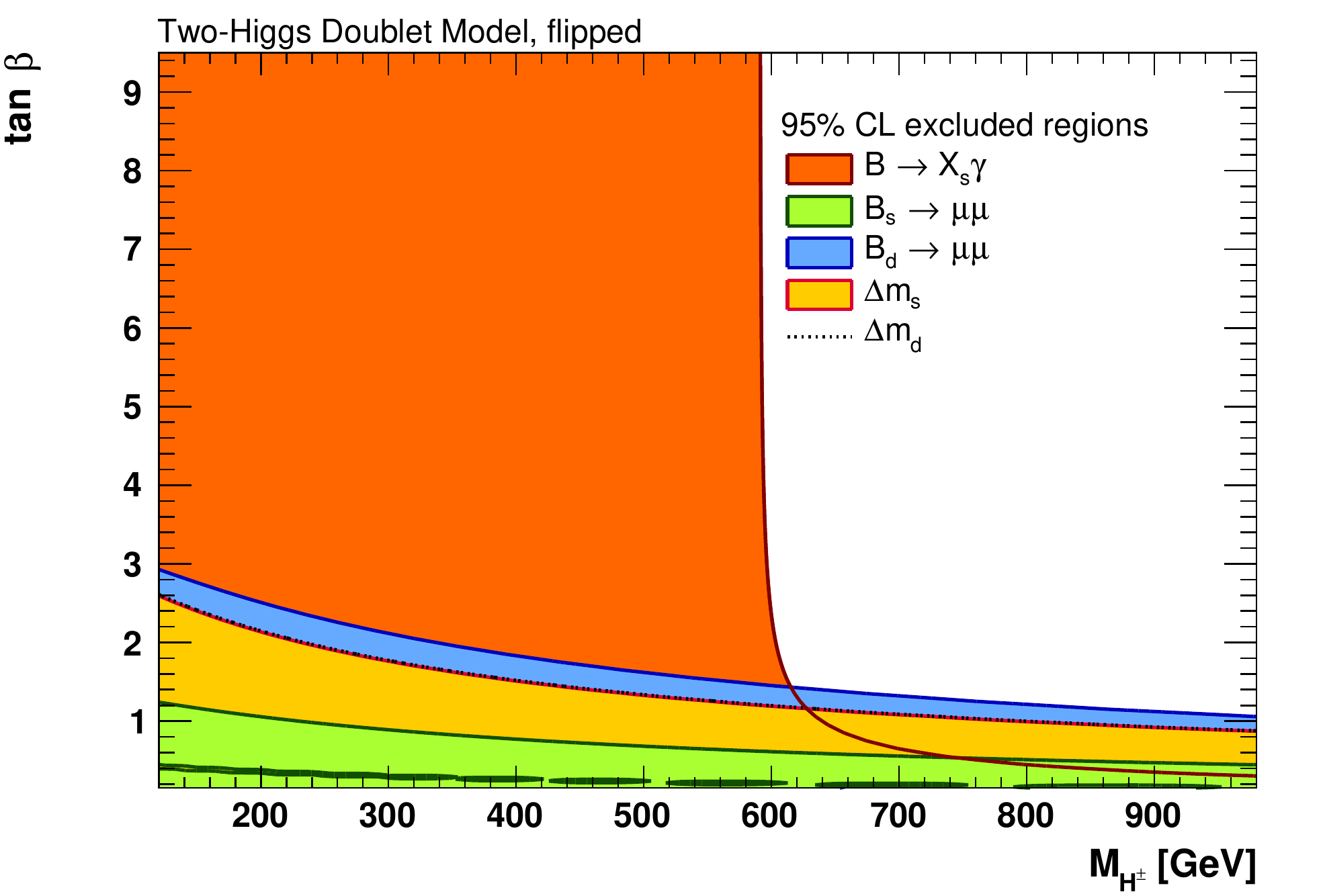}
\vspace{-0.8cm}
\end{center}
\caption[]{Excluded parameter regions (95\% CL) in the $\tanb$ versus $M_{H^\pm}$ plane 
  from individual observables for the four  2HDM scenarios considered: 
  Type-I (top left), Type-II (top right), lepton specific (bottom left), flipped (bottom right).}
\label{fig:TypeItoIV_flav}
\end{figure}
Figure~\ref{fig:TypeItoIV_flav} shows for the four 2HDM scenarios the one-sided 95\% CL excluded 
regions in the $\tanb$ versus $M_{H^\pm}$ plane as obtained from fits using the most sensitive individual 
flavour observables. The CLs are derived assuming a Gaussian behaviour of the test statistic with one 
degree of freedom. The Type-I (top left) and lepton specific (bottom left) scenarios are only weakly 
constrained allowing to exclude $\tanb<1$. Stronger constraints are obtained for the Type-II (top right) 
and flipped (bottom right) scenarios in which in particular $\BR(B\rightarrow X_s\gamma)$  allows to exclude 
$M_{H^\pm}<590\;$GeV.\footnote{Our results are compatible with those of Ref.~\cite{Misiak:2017bgg}, 
where limits on $M_{H^\pm}$ between 570 and 800\;GeV are reported for the Type-II model,
depending on the statistical method used (the CL has a relatively weak gradient
versus $M_{H^\pm}$ and thus exhibits a strong numerical sensitivity to the details of the interpretation). 
Similar exclusion limits on $M_{H^\pm}$ can be achieved in a 
complex 2HDM (C2HDM), which features additional mixing between the neutral CP-even and CP-odd
Higgs bosons~\cite{Muhlleitner:2017dkd}.}

The measurements of $R(D)$ and $R(D^*)$ differ from their SM 
predictions~\cite{Aoki:2016frl,Fajfer:2012vx, Na:2015kha}. 
In the 2HDM only the Type-II scenario features a compatible parameter 
region (at large $\tanb$ and relatively 
small $M_{H^\pm}$, not shown in the upper right plot of Fig.~\ref{fig:TypeItoIV_flav}), which is, 
however, excluded by several other  observables. Similar results have been reported 
in Ref.~\cite{Enomoto:2015wbn}.  Because of this incompatibility $R(D)$ and $R(D^{(*)})$ 
are excluded from our analysis in the following.

\subsection{Constraints from the anomalous magnetic moment of the muon}

The measured value of the anomalous magnetic moment of the muon 
$a_\mu=(g_\mu-2)/2$ shows a long-standing tension with the SM prediction 
of $\Delta a_\mu=(268\pm63\pm43)\cdot10^{-11}$~\cite{Bennett:2006fi,Davier:2017zfy},
where the first uncertainty is due the the measurement and the second the prediction
(see also the recent reanalysis in Ref.~\cite{Keshavarzi:2018mgv}). Loops involving 2HDM 
bosons can modify the coupling between photons and muons. We have adopted the two-loop
2HDM prediction of $\Delta a_\mu$  from Ref.~\cite{Cherchiglia:2016eui}, which 
depends on all 2HDM parameters. We make use of the code implementation kindly provided 
by H.~St\"ockinger-Kim.

\begin{figure}[t]
\begin{center}
\includegraphics[width=\defaultDoubleFigureScale\textwidth]{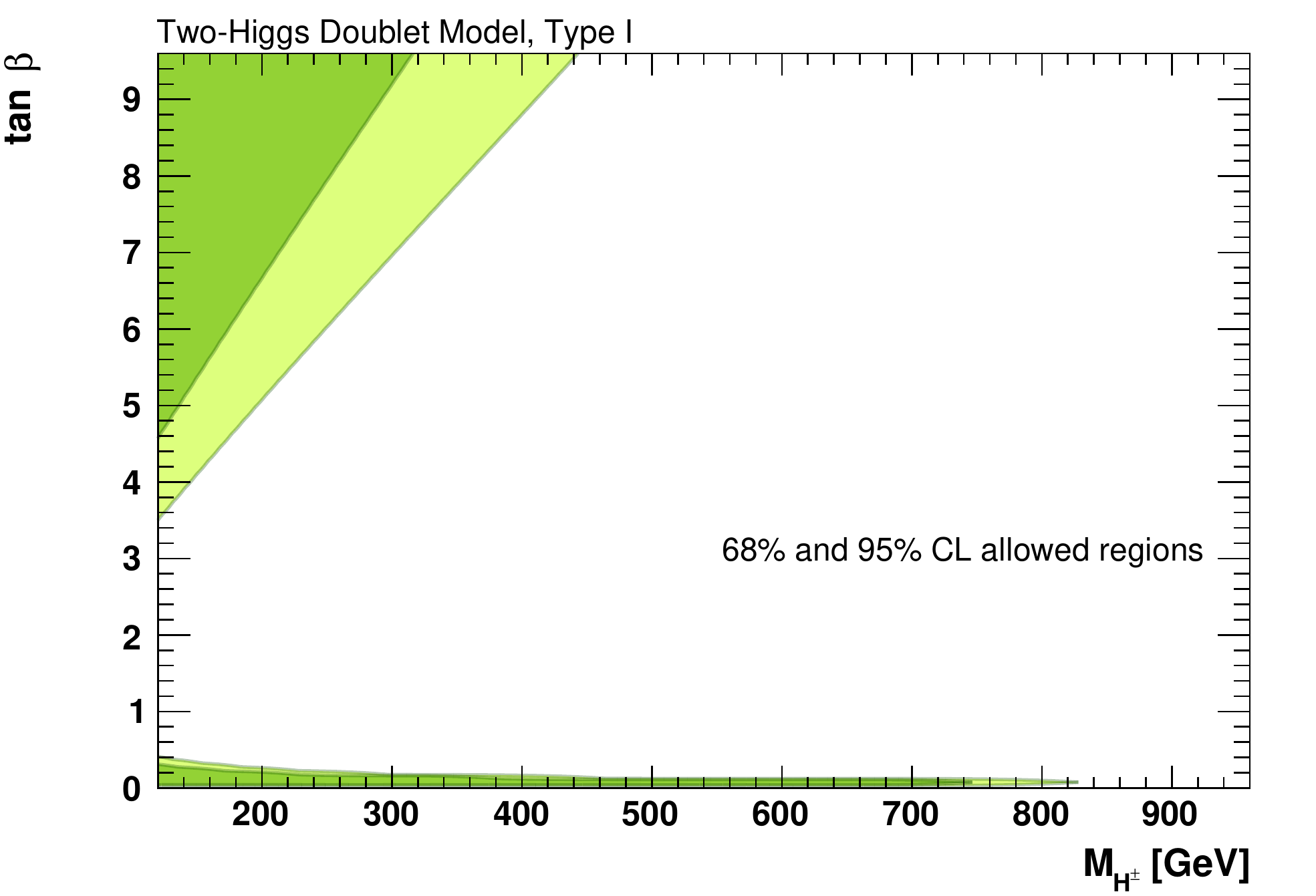}
\includegraphics[width=\defaultDoubleFigureScale\textwidth]{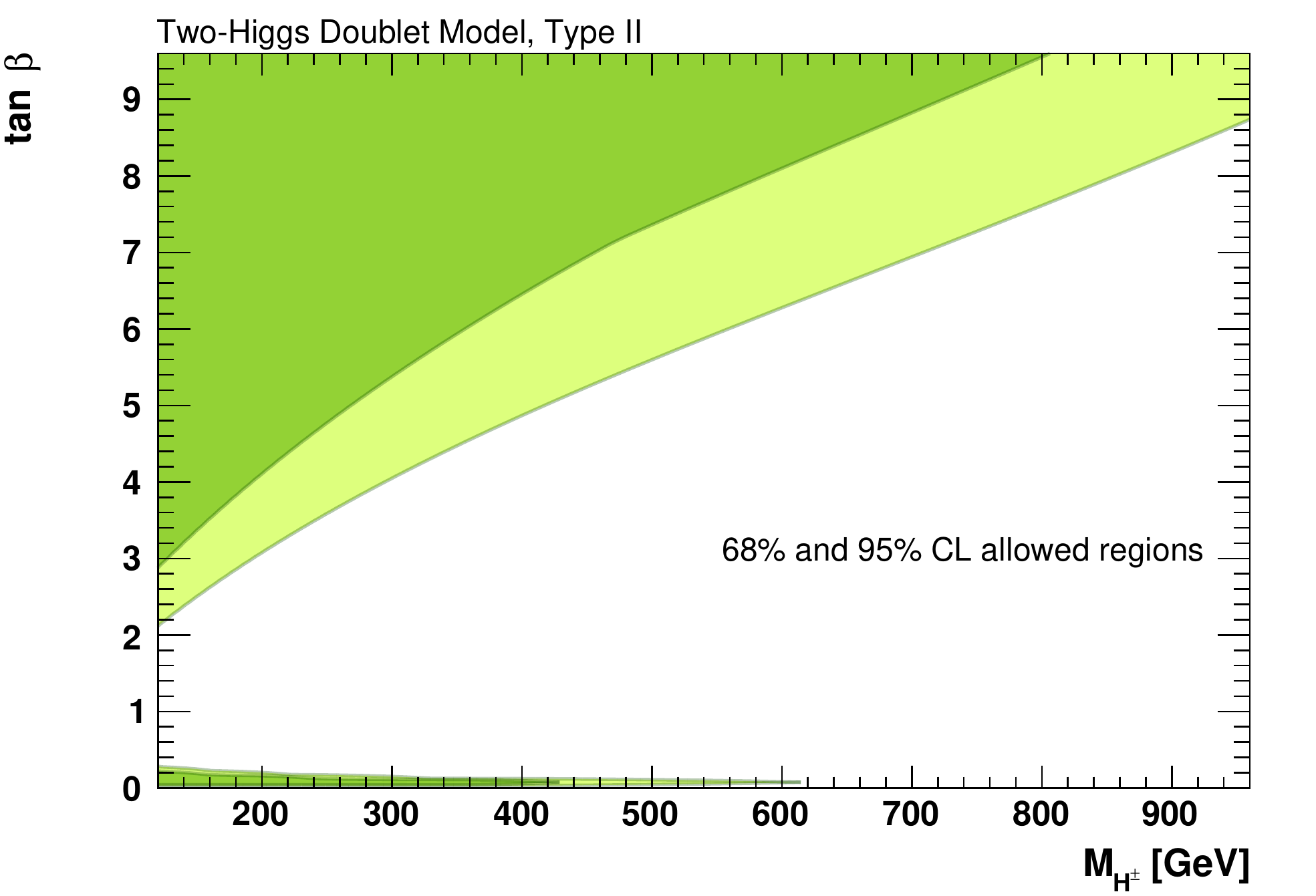}
\includegraphics[width=\defaultDoubleFigureScale\textwidth]{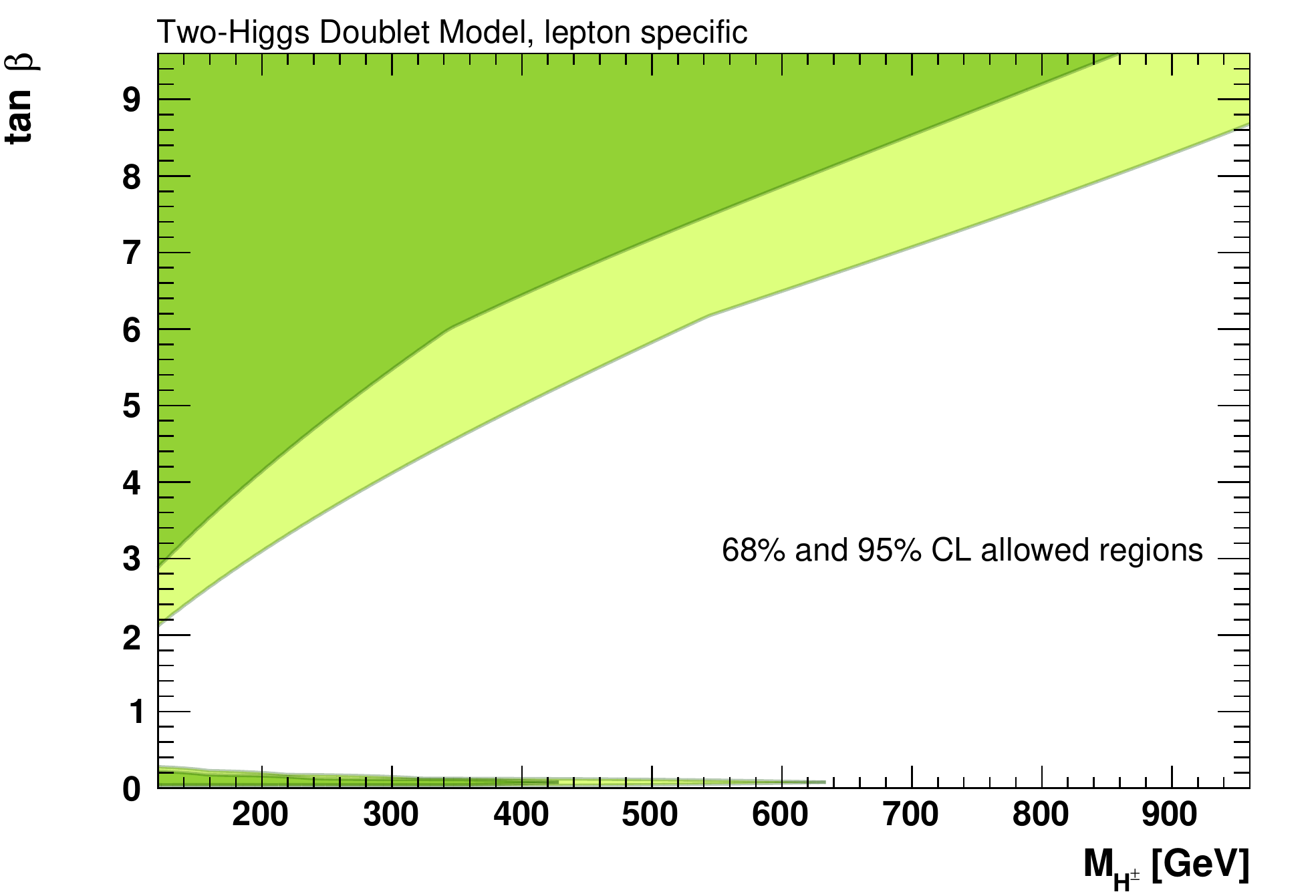}
\includegraphics[width=\defaultDoubleFigureScale\textwidth]{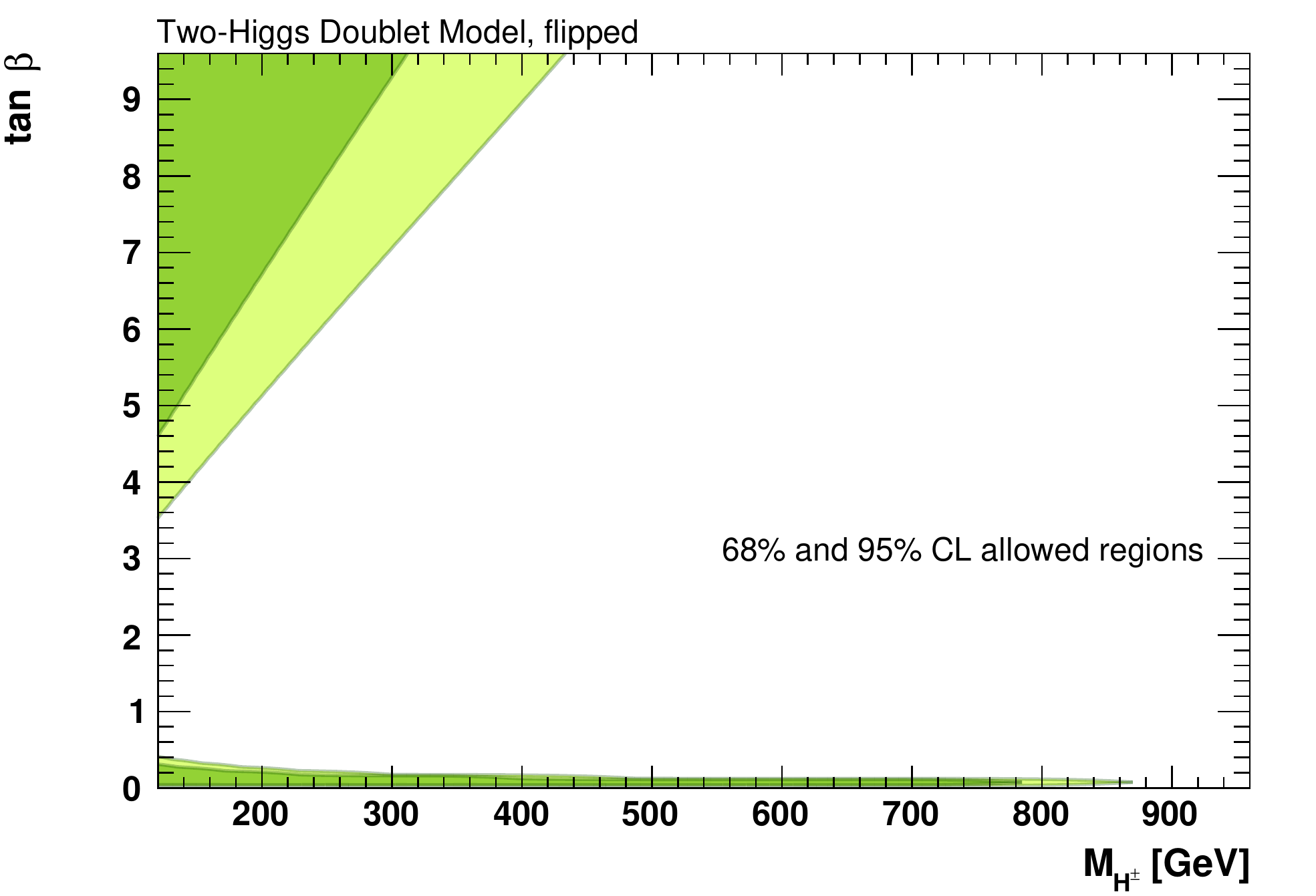}
\vspace{-0.8cm}
\end{center}
\caption[]{2HDM fits using the anomalous magnetic moment of the muon as input. Shown are 
  allowed 68\% and 95\% CL regions in the $\tanb$ versus $M_{H^\pm}$ plane for the four 
  2HDM scenarios considered: Type-I (top left), Type-II (top right), lepton specific (bottom left), 
  and flipped (bottom right).}
\label{fig:amu}
\end{figure}
Figure~\ref{fig:amu} shows the 68\% and 95\% CL allowed regions
in the $\tanb$ versus $M_{H^\pm}$ plane for the four 2HDM
scenarios using only $\Delta a_\mu$ as input.
All other parameters of the 2HDM  are left free to vary within their respective bounds. 
Compatibility is found in a narrow band with $\tanb\ll1$ and 
$M_{H^\pm}$ below about 600\;GeV (depending on the scenario), 
as well as for a region with larger $\tanb$ that broadens with 
decreasing $M_{H^\pm}$.  When combined with the constraints from 
the other flavour observables (cf. Fig.~\ref{fig:TypeItoIV_flav}),  
values of $\tanb$ above about 5$\sim$10 remain allowed.

\subsection{Constraints from electroweak precision data}

The electroweak precision data can be used to constrain the 2HDM via the 
oblique parameters determined in Eq.~(\ref{eq:stu}). We use the predictions from 
Refs.~\cite{Haber:1993wf,Haber:1999zh,Froggatt:1991qw} similar to our previous analysis~\cite{Baak:2011ze}.
The oblique corrections to electroweak observables in the 2HDM are independent of the Yukawa 
interactions and their impact is identical in the four 2HDM scenarios considered.

\begin{figure}
\begin{center}
\includegraphics[width=\defaultSingleFigureScale\textwidth]{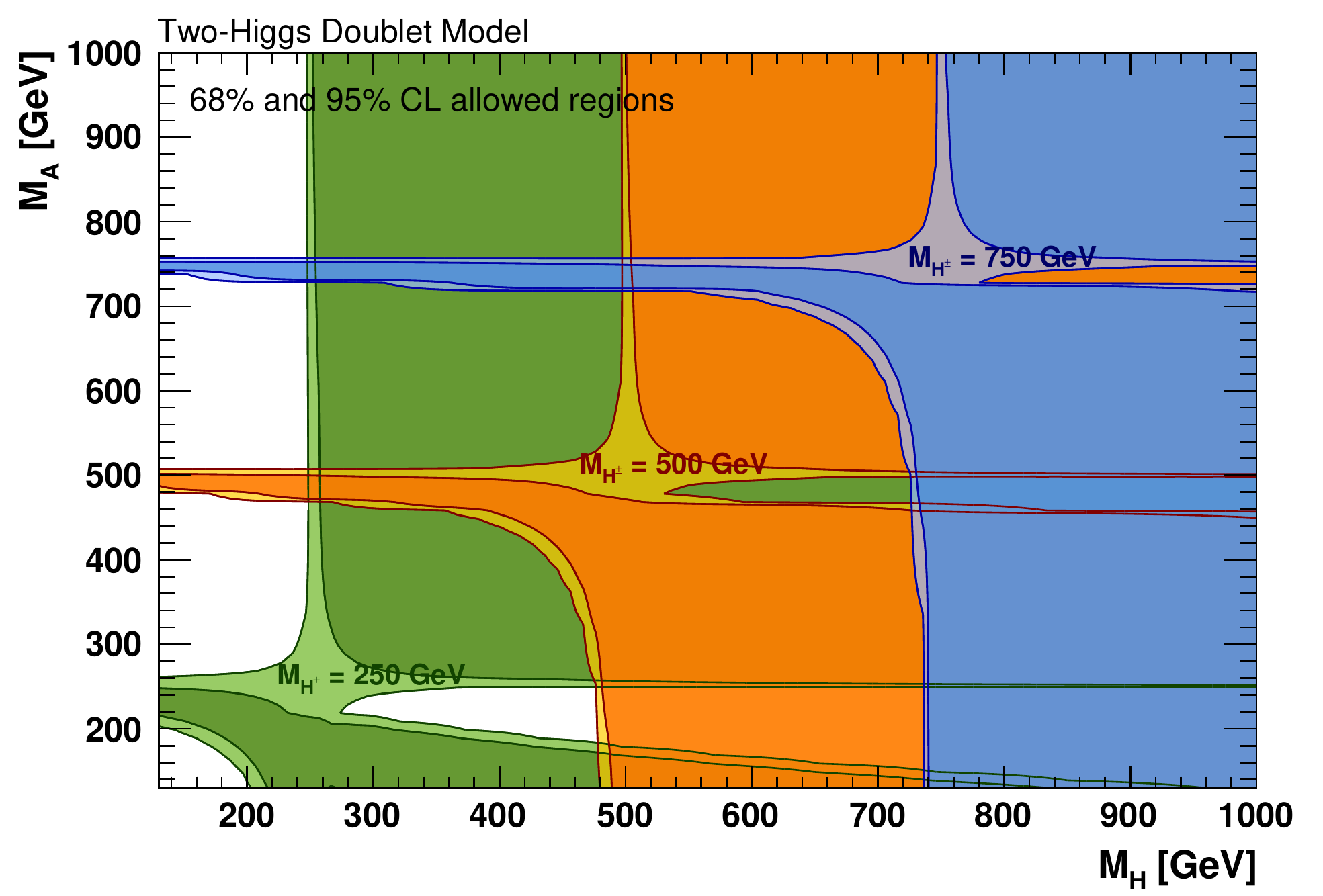}
\end{center}
\vspace{-0.3cm}
\caption[]{2HDM fit results using the oblique $S$, $T$, $U$ parameters.
  Shown are allowed 68\% and 95\% CL   regions in the  $M_{A}$ versus $M_{H}$ 
  plane for fixed benchmark values of $M_{H^{\pm}}$. 
  The constraints are independent of the 2HDM scenario.}
\label{fig:2HDM_STU}
\end{figure}

Figure~\ref{fig:2HDM_STU} shows the 68\% and 95\% CL allowed parameter
regions in the neutral Higgs-boson mass plane $M_{A}$ versus $M_{H}$
for fixed charged Higgs-boson masses of 250, 500, and
750\;GeV as obtained from fits using only the oblique parameters as input.  All
other parameters of the 2HDM (including $\beta-\alpha$) are free
to vary in these scans.  While no information on the absolute mass scale of 
the 2HDM bosons is obtained from the electroweak data, relative masses
are constrained. In our previous
analysis~\cite{Baak:2011ze} we showed that the oblique
parameters constrain the values of $M_{H}$ and $M_{A}$ to be close
to $M_{H^{\pm}}$ for fixed  $\beta-\alpha=\pi/2$. Removing this  
restriction (cf. Fig.~\ref{fig:2HDM_STU}) relaxes 
the constraint to having either  $M_{A}$ close to $M_{H^{\pm}}$,
or $M_{H}$ larger than $M_{H^{\pm}}$.

\subsection{Combined fit}
\label{sec:combi}

We combine in this section the 2HDM constraints from the Higgs-boson coupling strength 
measurements,  flavour observables, muon anomalous magnetic moment, and electroweak precision data.

\begin{figure}[t]
\begin{center}
\includegraphics[width=\defaultDoubleFigureScale\textwidth]{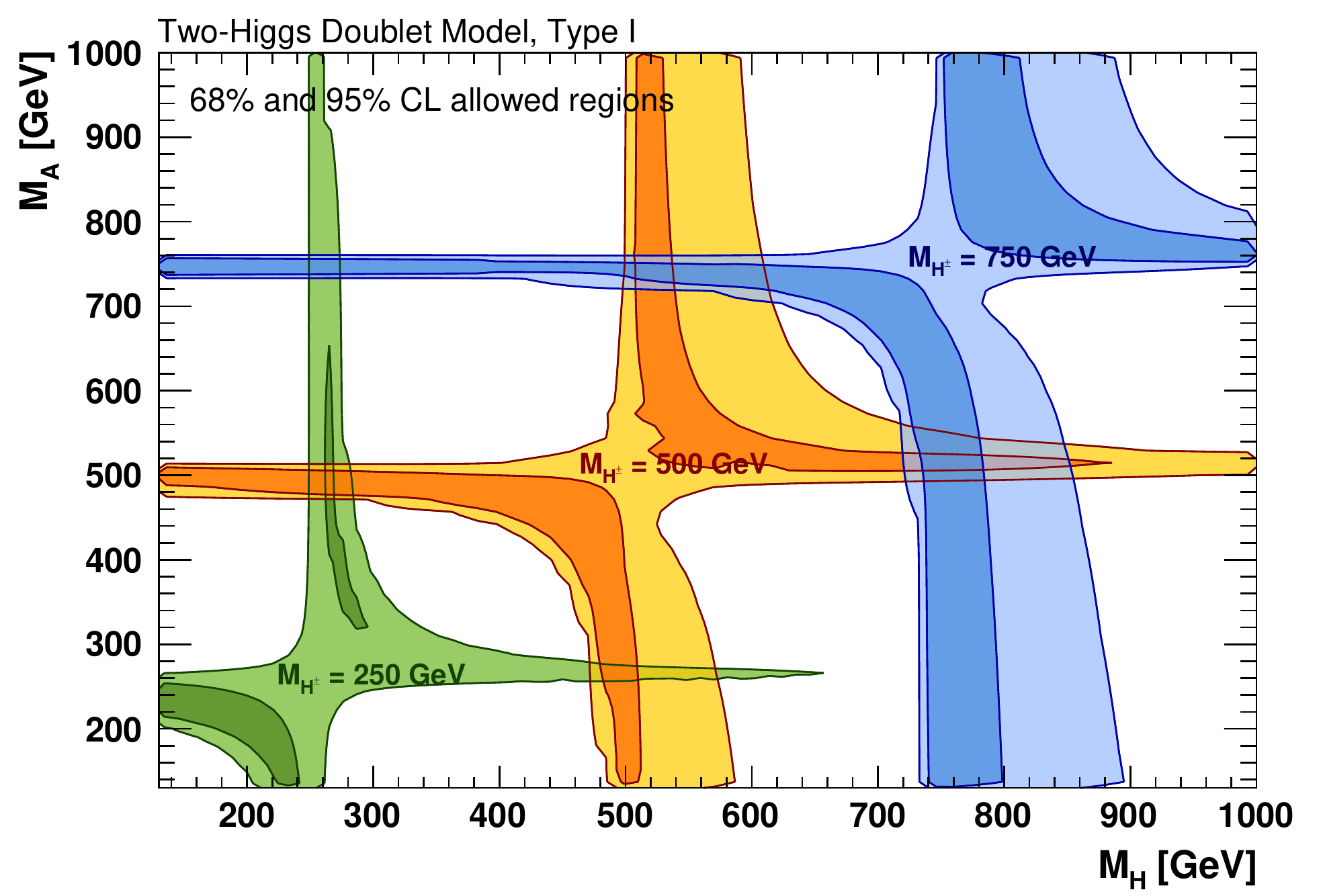}
\includegraphics[width=\defaultDoubleFigureScale\textwidth]{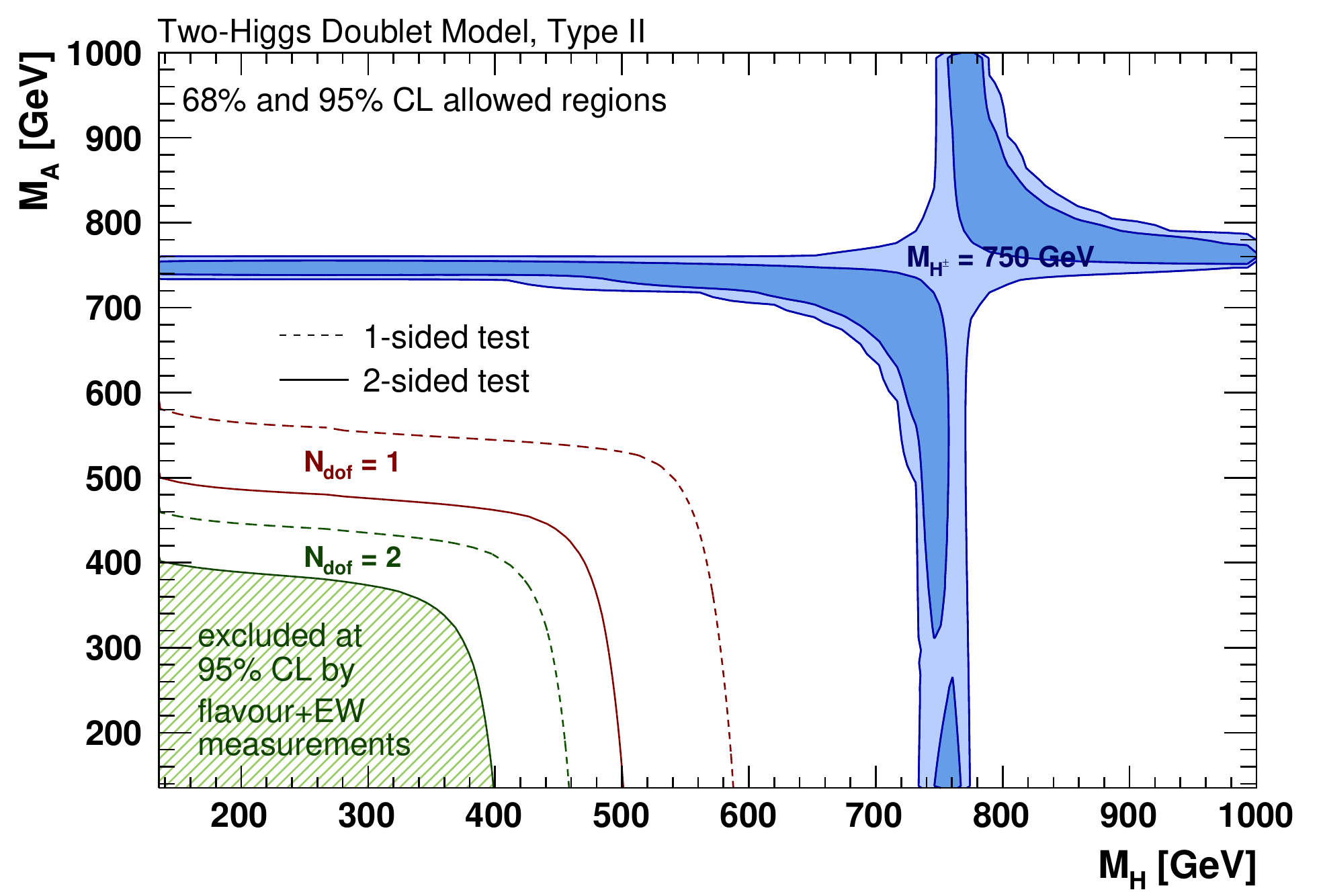}
\includegraphics[width=\defaultDoubleFigureScale\textwidth]{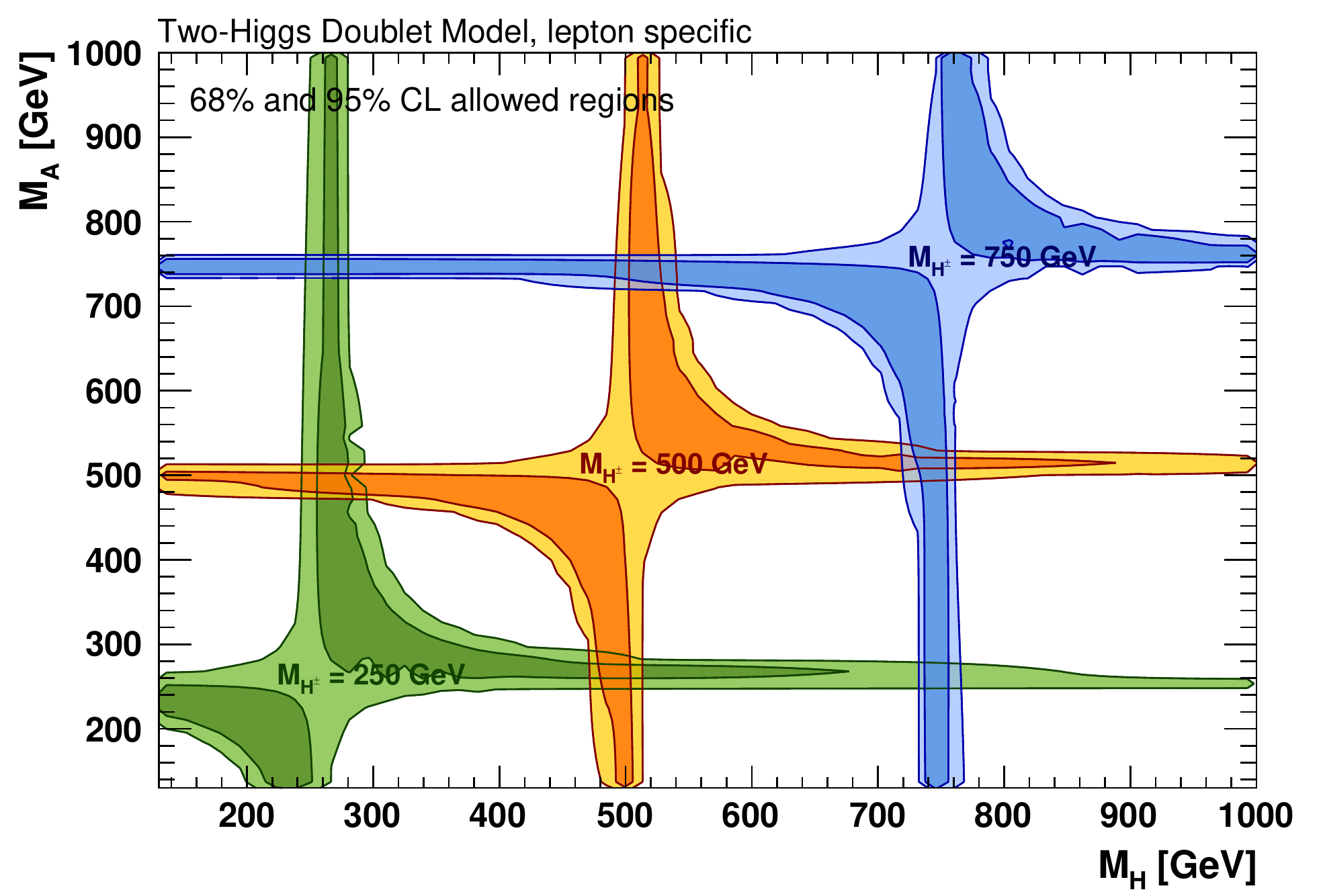}
\includegraphics[width=\defaultDoubleFigureScale\textwidth]{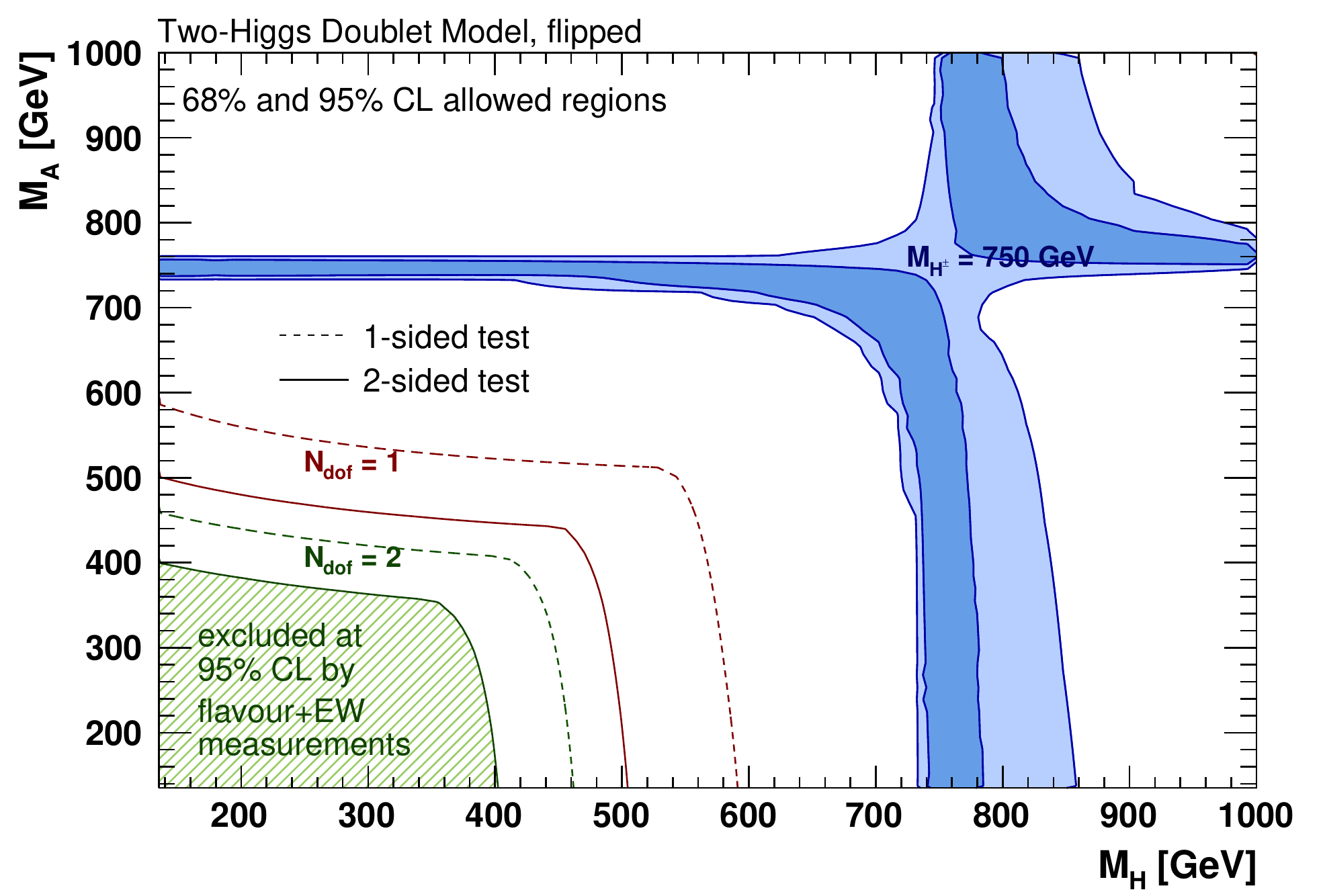}
\vspace{-0.8cm}
\end{center}
\caption[]{2HDM fit results using a combination of constraints from the Higgs-boson coupling strength 
  measurements,  flavour observables, muon anomalous magnetic moment, and electroweak precision data.
  Shown are allowed 68\% and 95\% CL  regions in the $M_{A}$ versus $M_{H}$ plane  for fixed benchmark
  values of $M_{H^{\pm}}$ and for the four  2HDM scenarios considered: Type-I (top left), Type-II (top right), 
  lepton specific (bottom left), and flipped (bottom right).}
\label{fig:MA_vs_MH0}
\end{figure}

Figure~\ref{fig:MA_vs_MH0} shows for the four 2HDM scenarios considered 
the resulting 68\% and 95 \% CL allowed regions in the $M_{A}$ 
versus $M_{H}$ plane for fixed (benchmark) charged Higgs-boson masses of 250, 500, and 750\;GeV. 
All other 2HDM parameters are allowed to vary freely within their bounds. 
Depending on the 2HDM scenario and $M_{H^{\pm}}$, the minimum $\chi^2$ values found
lie between 48 and 59 for $\Ndof=53$ (corresponding to $p$-values between 25\% and 68\%).

The combined fit leads in all four 2HDM scenarios to a strong alignment of either the $H$ or the 
$A$ boson mass with that of the $H^{\pm}$ boson, owing to the constraint on $\beta-\alpha$ from
the Higgs coupling strength measurements (cf. Fig.~\ref{fig:bma_tanb}) in addition to those
from the electroweak precision data. 
In this sense, the fit resembles the result from our previous analysis~\cite{Baak:2011ze},
but replacing the fixed restriction of $\beta-\alpha=\pi/2$ by the Higgs couplings strengths measurements.

The absolute mass limits on $M_{\rm H^{\pm}}$ obtained from the flavour observables in the Type-II 
and flipped scenarios (cf. Fig.~\ref{fig:TypeItoIV_flav}) exclude the low-$M_{\rm H^{\pm}}$ benchmarks, 
as indicated by the hatched regions in the two right-hand panels of Fig.~\ref{fig:MA_vs_MH0} 
(where in addition different statistical assumptions are compared: one-sided versus two-sided test statistic and 
one versus two degrees of freedom\footnote{The limits obtained for a two-sided test statistic and two 
degrees of freedom have been verified with a pseudo Monte Carlo study based on randomly 
drawn sets of the measurements used in the fit.}). For these two scenarios pairs of ($H$, $A$) masses below 
$\sim$$400\;\gev$ are excluded for any set of values of the other 2HDM parameters.
For the Type-I and lepton specific scenarios no absolute limits on the Higgs boson masses 
can be derived.

\section{Conclusion}
\label{sec:conclusion}

We have presented results for an updated global fit of the electroweak
sector of the Standard Model using  latest experimental and theoretical input. 
We include new precise kinematic top quark and 
$W$ boson mass measurements from the LHC, a \sinleff measurement
from the Tevatron, and a new evaluation of the hadronic contribution
to $\alpha(M_Z^2)$. 
The fit confirms the consistency of the Standard Model and slightly improves
the precision of the indirect determination of key observables.

Using constraints from  Higgs-boson coupling strength measurements,  flavour 
observables, the muon anomalous magnetic moment, and electroweak precision data,
we studied allowed and excluded parameter regions of four CP conserving two-Higgs-doublet 
models. Strong constraints on the extended Higgs boson masses 
are found for the so-called Type-II and flipped scenarios.

\subsubsection*{Acknowledgements}
\label{sec:Acknowledgments}

\begin{details}
  We are indebted to Mikolaj Misiak and Hyejung St\"ockinger-Kim for providing the implementation
  of their calculations of $\BR(B\rightarrow X_s\gamma)$ and the muon anomalous magnetic moment 
  in the two-Higgs-doublet models.
  We  thank Rui Santos for helpful discussions and feedback on early stages of the paper.
  This work is supported by  the German Research Foundation (DFG) in the Collaborative Research
  Centre (SFB) 676 ``Particles, Strings and the Early Universe''  located in Hamburg.

\appendix

\section*{Appendix}
\label{appendix}

To validate our implementation of the Higgs boson coupling  measurements with respect
to the full result from the ATLAS and CMS combination~\cite{Khachatryan:2016vau}, we have 
performed a fit of a generic new physics parametrisation. Here, new physics effects are assumed 
to uniformly vary the coupling strength of the Higgs boson to vector bosons and fermions, respectively, 
according to linear modifiers $\kappa_V$ and $\kappa_F$.
No new particles are assumed to contribute to the Higgs boson production via loop diagrams and 
the branching fraction of the Higgs boson to unknown  states is assumed to be zero.
The constraints on $\kappa_V$ and $\kappa_F$ from the individual Higgs boson decay channels and 
their combination are shown in Fig.~\ref{fig:CVCF}. We obtain the best fit values 
$\kappa_V=1.00\pm0.05$ and $\kappa_F=0.92\pm0.11$ with a correlation coefficient of $-0.37$.
Decent agreement with Ref.~\cite{Khachatryan:2016vau} is seen.

\begin{figure}
\begin{center}
\includegraphics[width=\defaultSingleFigureScale\textwidth]{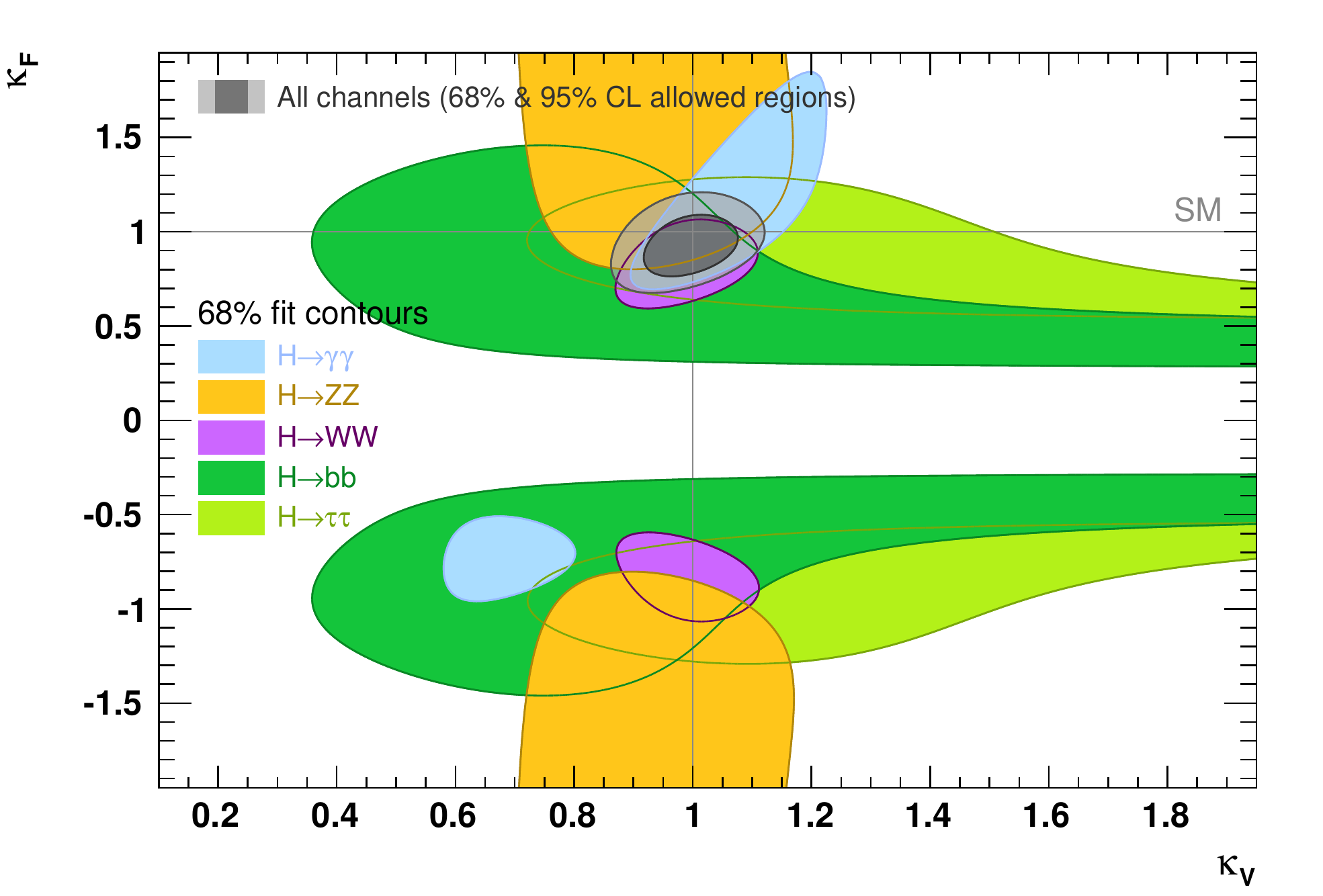}
\end{center}
\vspace{-0.3cm}
\caption[]{Validation of our implementation of the combined ATLAS and CMS 
  Higgs boson coupling measurements: 
  preferred regions from a two-dimensional scan of the coupling strength 
  modifiers $\kappa_V$ and $\kappa_F$ for individual Higgs boson decay channels
  and their combination. }
\label{fig:CVCF}
\end{figure}

\end{details}

\addcontentsline{toc}{section}{References}
\bibliography{References}{}

\end{document}